\documentclass[a4paper,fleqn,usenatbib,useAMS]{mnras}

\usepackage{graphicx}	
\usepackage{amsmath}	
\usepackage{amssymb}	
\usepackage{multicol}        
\usepackage{bm}		
\usepackage{pdflscape}	

\usepackage[T1]{fontenc}
\usepackage{ae,aecompl}

\usepackage{newtxtext,newtxmath}

\title[How to Quench a Dwarf Galaxy]{How to Quench a Dwarf Galaxy: The Impact of Inhomogeneous Reionization on Dwarf Galaxies and Cosmic Filaments}

\author[H. Katz et al.] {Harley Katz$^{1}$\thanks{E-mail:
  harley.katz@physics.ox.ac.uk},
  Marius Ramsoy$^1$
  Joakim Rosdahl$^{2}$,
  Taysun Kimm$^{3}$,
  J\'er\'emy Blaizot$^{2}$,
  \newauthor 
  Martin G. Haehnelt$^{4}$,
  L\'eo Michel-Dansac$^{2}$,
  Thibault Garel$^{5,2}$,
  Clotilde Laigle$^1$,
  \newauthor
  Julien Devriendt$^{1}$,
  and Adrianne Slyz$^{1}$. \\ \\
  $^1$Sub-department of Astrophysics, University of Oxford,
  Keble Road, Oxford OX1 3RH, UK \\
  $^2$Univ Lyon, Univ Lyon1, Ens de Lyon, CNRS, Centre de Recherche
  Astrophysique de Lyon UMR5574, F-69230, Saint-Genis-Laval, France \\
  $^3$Department of Astronomy, Yonsei University, 50 Yonsei-ro,
  Seodaemun-gu, Seoul 03722, Republic of Korea \\  
  $^4$Kavli Institute for Cosmology and Institute of Astronomy,
  Madingley Road, Cambridge CB3 0HA, UK \\
  $^5$Observatoire de Geneve, Universite de Geneve, 51 Ch. des Maillettes, 1290 Versoix, Switzerland\\
  }
\date{\today}

\pubyear{2017}

\begin{document}
\label{firstpage}
\pagerange{\pageref{firstpage}--\pageref{lastpage}}
\maketitle

\begin{abstract}
We use the {\small SPHINX} suite of high-resolution cosmological radiation hydrodynamics simulations to study how spatially and temporally inhomogeneous reionization impacts the baryonic content of dwarf galaxies and cosmic filaments.  The {\small SPHINX} simulations simultaneously model an inhomogeneous reionization, follow the escape of ionising radiation from thousands of galaxies, and resolve haloes well below the atomic cooling threshold. This makes them an ideal tool for examining how reionization impacts star formation and the gas content of dwarf galaxies.  We compare simulations with and without stellar radiation to isolate the effects of radiation feedback from that of supernova, cosmic expansion, and numerical resolution.  We find that the gas content of cosmic filaments can be reduced by more than 80\% following reionization.  The gas inflow rates into haloes with ${\rm M_{vir}\lesssim10^8M_{\odot}}$ are strongly affected and are reduced by more than an order of magnitude compared to the simulation without reionization.  A significant increase in gas outflow rates is found for halo masses ${\rm M_{vir}\lesssim7\times10^7M_{\odot}}$.  Our simulations show that inflow suppression (i.e. starvation), rather than photoevaporation, is the dominant mechanism by which the baryonic content of high-redshift dwarf galaxies is regulated.  At fixed redshift and halo mass, there is a large scatter in the halo baryon fractions that is entirely dictated by the timing of reionization in the local region surrounding a halo which can change by $\Delta z\gtrsim3$ at fixed mass.  Finally, although the gas content of high-redshift dwarf galaxies is significantly impacted by reionization, we find that most haloes with ${\rm M_{vir}\lesssim10^8M_{\odot}}$ can remain self-shielded and form stars long after reionization, until their local gas reservoir is depleted, suggesting that local group dwarf galaxies do not necessarily exhibit star formation histories that peak prior to $z=6$. Significantly larger simulation boxes will be required to capture the full process of reionization and understand how our results translate to environments not probed by our current work.

\end{abstract}




\section{Introduction}
Galaxy formation in the lowest mass dark matter haloes is observed to be extremely inefficient \citep{McGaugh2010}.  Likewsie, certain ultra-faint dwarf galaxies around our own Milky Way exhibit extremely old stellar populations, consistent with most stars forming at $z>6$ \citep{Brown2014}.  This suggests that some process may be acting to shut down star formation in the lowest mass galaxies in the early Universe.  Furthermore, such a process may help explain, the inefficiency of star formation, the old stellar populations, and relieve tension in some of the small-scale problems of $\Lambda$CDM including the missing satellites problem \citep{Klypin1999,Moore1999}, the too-big-to-fail problem \citep{Boylan2011}, and the core/cusp problem \citep{Moore1994,Flores1994}.

Haloes with circular velocities $V_{\rm c}\lesssim30{\rm km\ s^{-1}}$ are particularly susceptible to various forms of energetic feedback including that from supernova \citep[SN, e.g.][]{Dekel1986,MC1999} and UV radiation \citep[e.g.][]{Rees1986,Bullock2000}.  During the first billion years of cosmic evolution, the first generations of stars, galaxies, and black holes emit copious amounts of ionising photons that form the meta-galactic UV background and reionize the Universe.  This process is expected to suppress the formation of dwarf galaxies via photoevaporation \citep{Rees1986,Shapiro2004,Iliev2005}, by prolonging cooling times \citep{Efstathiou1992}, and by preventing gas inflows \citep{Shapiro1994,Gnedin2000b,Hoeft2006,Okamoto2008,Gnedin2014}.  

Various observational campaigns have attempted to measure the star formation histories of local group dwarf galaxies in order to identify whether there is direct evidence for a drastic decrease in the star formation rates of these systems at $z=6$ that can be associated with reionization.  Currently there is no general consensus. Certain observations indicate that there are some galaxies that exhibit features in their star formation histories consistent with suppression due to reionization \citep{Weisz2014,Brown2014,Skillman2017,Bett2018}, while various others show no definite signature \citep{Grebel2004,Monelli2010,Hidalgo2011,Weisz2014,Skillman2017}.  The limited time resolution of the inferred star formation histories in the early Universe from colour-magnitude diagrams makes these observations particularly difficult \citep{Aparicio2016}.  Furthermore, new theoretical work indicates that star formation may reignite long after reionization \citep[e.g.][]{Benitez2015,Wright2018,Ledin2018} which complicates the interpretation of the role of reionization on dwarf galaxy suppression from inferred star formation histories.  

The effects of reionization on the gas content of low mass dwarf galaxies are well established by high-resolution numerical simulations, independent of whether radiation transfer is self-consistently modelled in the simulation, or an approximate UV background is used \citep[e.g.][]{Gnedin2000b,Hoeft2006,Okamoto2008,Wise2014,Gnedin2014,Sawala2016,Oshea2015,Ocvirk2016,Xu2016,Ma2017,Fitts2017,Wu2019}.  Nevertheless, disagreement among simulations still persists regarding the characteristic mass at which dwarf galaxy formation is suppressed, the timing and environmental dependence of the suppression, the effect on the subsequent star formation rates, and the relative importance of the different modes of suppression (i.e., photoevaporation versus preventing gas inflows).  Certain aspects of these open questions, such as the environmental dependence of the suppression, can only be readily addressed in simulations that couple an inhomogeneous and time-dependent radiation field to the gas dynamics and chemistry in a cosmological simulation.  Strong fluctuations in the strength of the UV background are expected during reionization as well as in the post-reionization epoch \citep{Becker2015,Bosman2018,Kulkarni2018} and this may lead to different amounts of photoevaporation depending on environment.  Furthermore, the timing of reionization changes depending on environment which can also impact both the amount of photoevaporation and reduction of gas inflows \citep{Sobacchi2013}.  

Understanding the preventative feedback that inhibits gas accretion onto dwarf galaxies appears key for explaining numerous characteristics of Milky Way and Local Group dwarf galaxies \citep{Lu2017}.  \cite{Noh2014} attempted to develop a physical model for how reionization suppresses gas accretion by comparing scales of density, cooling, photoheating, and self-shielding, by following the complex trajectories of parcels of gas traversing temperature-density phase-space.  They find that gas will likely accrete onto a halo as long as the halo is massive enough so that the gas is Jeans unstable along its entire trajectory in temperature-density phase-space.  This model focuses primarily on the post-reionization epoch and on spherical accretion whereas early on, galaxies may be primarily fed via thin, dense, cold filaments \citep[e.g.][]{Keres2005}.  

It is already established that the width of gas filaments increases due to the increase in thermal pressure  when the filaments become photoionised during reionization \citep{Theuns2000,Pawlik2009,Wise2014,Ocvirk2016}.  \cite{Ocvirk2016} extracted the properties of a single, massive gas filament from their cosmological radiation hydrodynamics simulations and demonstrated that the filament is surrounded by a casing of hot gas, above the background temperature, while containing a cool central core.  The general picture of accretion onto filaments is similar to that of galaxies whereby gas is expected to shock heat at the edge of the filament, causing it to stall, become denser, and cool as it is gravitationally pulled to the centre.  Note that filaments are only partially gravitationally bound along two dimensions and are not expected to be self-shielded well after reionization \citep[e.g.][]{Finlator2009}.  The filament studied in \cite{Ocvirk2016} was relatively massive and thus had a shock temperature of $\sim10^5$K, which is much greater than the temperature of $\sim10^4$K to which gas is heated to during reionization \citep[e.g.][]{Puchwein2018}.  We expect that the filaments that are likely to be most affected by reionization are those with shock temperatures $\lesssim10^4$K which are correspondingly expected to feed the least massive galaxies.  If the gas can no longer shock at the edge of the filament, it is unlikely to cool efficiently to the centre or be retained at all, thereby reducing any potential accretion onto a dwarf galaxy that resides at a node fed by such a filament.

In this work, we employ the state-of-the-art {\small SPHINX} suite of cosmological radiation hydrodynamics simulations \citep{Rosdahl2018} to better characterise the effects of inhomogeneous reionization on the formation of dwarf galaxies and cosmic filaments.  These simulations have both the mass and spatial resolution needed to analyse the impact of radiation feedback on the smallest cosmic structures and resolve galaxies below the atomic cooling threshold.  Our work uses the highest resolution simulations in the {\small SPHINX} suite which are needed to properly resolve haloes at and below the atomic cooling threshold; however, this comes at the expense of limited box size which impacts our ability to translate our results to a wide variety of environments that have drastically different reionization histories and photoionisation rates. We perform a systematic study on how the gas content of cosmic filaments is affected by reionization and how this directly relates to the gas accretion rates, gas content, and stellar content of dwarf galaxies during the first billion years.  Furthermore, we explore how the patchiness of reionization and environmental effects, such as local reionization redshift and thermodynamic state, impact the formation of low-mass haloes.  

This paper is organised as follows:  in Section~\ref{simdis}, we describe the {\small SPHINX} suite of cosmological radiation hydrodynamics simulations and our methods to extract galaxy, halo, and filament properties.  In Section~\ref{results}, we present our results on how reionization affects cosmic filaments and dwarf galaxies and elucidate the physics that is responsible for suppressing the formation of low-mass galaxies.  In Section~\ref{discon}, we present our discussion and conclusions.

\section{Numerical Methods}
In this Section, we outline the details of our simulation as well as our methods for extracting filament and halo properties.

\label{simdis}
\subsection{Cosmological Radiation Hydrodynamics Simulations}
This work makes use of the {\small SPHINX} simulations \citep{Rosdahl2018}, a suite of cosmological, multifrequency, radiation-hydrodynamics, adaptive mesh refinement simulations that models an inhomogeneous reionization and resolves the interstellar medium (ISM) within galaxies.  The simulations have been run with {\small RAMSES-RT} \citep{Rosdahl2013}, an extension of the open-source {\small RAMSES} code \citep{Teyssier2002}, that includes on-the-fly radiation transfer and six-species (H, H$^+$, e$^-$, He, He$^+$, H$^{++}$) non-equilibrium chemistry.  The full details of the simulation suite can be found in Section~2 of \cite{Rosdahl2018} and here, we briefly review the basic details of the simulations.

The hydrodynamics is solved using an HLLC Riemann solver \citep{Toro1994} and a MinMod slope limiter.  We assume an adiabatic index of $\gamma=5/3$ to close the relation between gas pressure and internal energy.  Gravity is computed using a multigrid solver \citep{Guillet2011} and particles are projected onto the grid using cloud-in-cell interpolation.  The radiation is advected between cells using a first-order moment method, using the M1 closure \citep{Levermore1984} and the Global-Lax-Friedrich intercell flux function.

The initial conditions of the simulations are generated with {\small MUSIC} \citep{Hahn2011} using the cosmological parameters $\Omega_{\Lambda}=0.6825$, $\Omega_{\rm m}=0.3175$, $\Omega_{\rm b}=0.049$, $h=0.6711$, and $\sigma_8=0.83$ \citep{Planck}.  The box is filled with a primordial mixture of neutral hydrogen and helium with mass fractions $X=0.76$ and $Y=0.24$, respectively.  A very small initial metallicity of $3.2\times10^{-4}Z_{\odot}$ (assuming $Z_{\odot}=0.02$) is assumed for all cells to compensate for the lack of molecular hydrogen cooling in our simulation.  This is calibrated so that star formation begins at $z\sim20$.

The {\small SPHINX} suite contains simulations with various volumes and spatial and mass resolutions.  For this work, we employ only the highest resolution simulations with a comoving box size of $5^3$Mpc$^3$, a maximum spatial resolution of 10.9pc at $z=6$ (the minimum cell width increases linearly with the expansion factor), and a dark matter particle mass resolution of $3.1\times10^4$M$_{\odot}$ (corresponding to $512^3$ particles).  This mass and spatial resolution is crucial in order to properly model the effect of UV radiation on low mass galaxies \citep[e.g.][]{Okamoto2008}.  The volumes have been selected from a set of 60 initial conditions in order to minimise the effect of cosmic variance as our box size is not large enough to capture the cosmological homogeneity scale (see Figure~1 of \citealt{Rosdahl2018}). 

Gas cooling follows the prescription presented in the Appendix of \cite{Rosdahl2013}.  Cooling from primordial species via collisional ionisation, recombination, collisional excitation, Bremsstrahlung, Compton cooling (and heating) from the cosmic microwave background, and di-electronic recombination are all included.  Cooling from metal lines is calculated by interpolating tables generated with {\small CLOUDY} \citep{Ferland1998} for temperatures $>10^4$K.  At temperatures below this value, fine structure line cooling rates are computed using the fits from \cite{Rosen1995}.

\begin{figure}
\centerline{\includegraphics[scale=1.0,trim={0cm 0.5cm 0cm 1cm},clip]{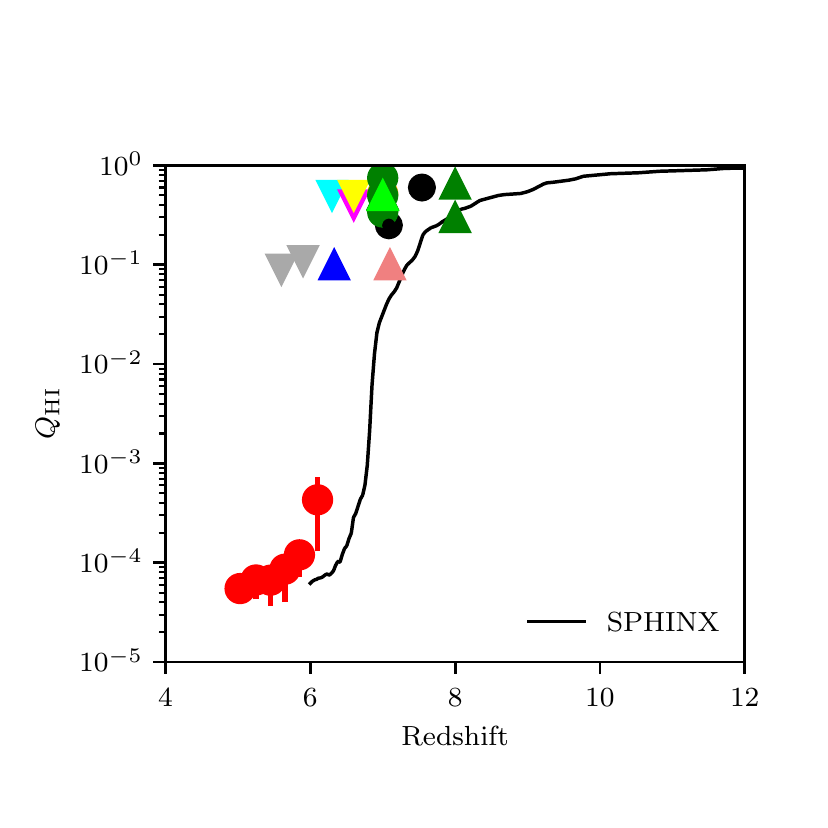}}
\centerline{\includegraphics[scale=1.0,trim={0cm 0.5cm 0cm 1cm},clip]{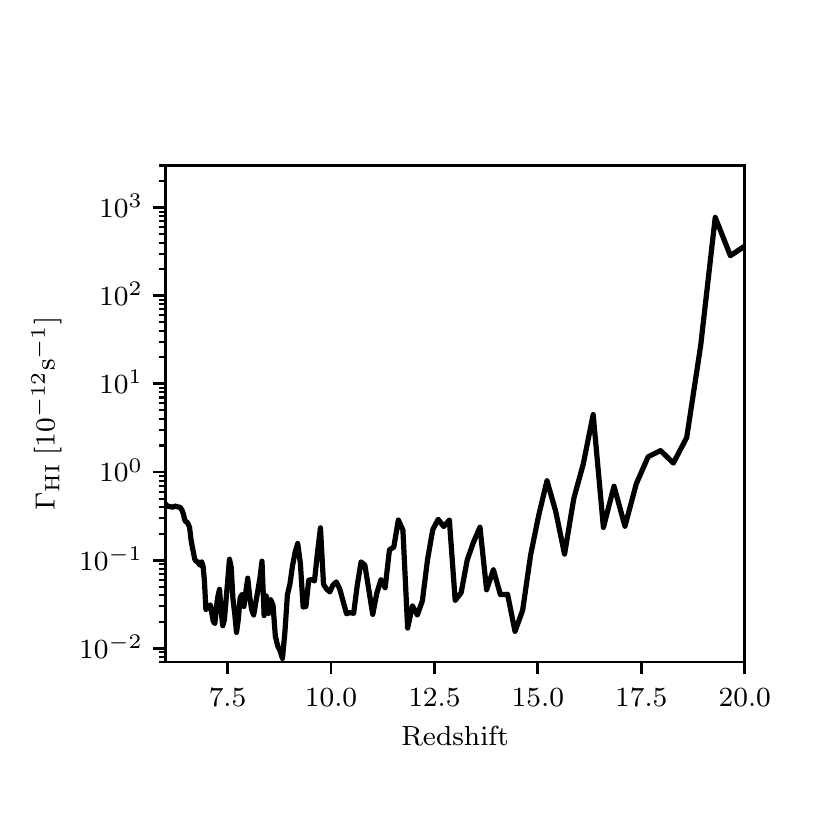}}
\caption{({\bf Top}) Reionization history of our simulation (black line) compared to observations (coloured points). $Q_{\rm HI}$ represents the volume weighted neutral fraction.  Observational data was compiled in \protect\cite{Bouwens2015} (see their Table~1 for references), excluding the black circles which are recent high-redshift constraints from \protect\cite{Durov2019}.  ({\bf Bottom}) Volume weighted HI photoionisation rate in ionised regions as a function of redshift.}
\label{QHI}
\end{figure}

Stellar radiation is modelled using three energy bins with lower energy limits of 13.60eV, 24.59eV, and 54.42eV, respectively, to capture the ionisation energies of H, He, and He$^+$.  The on-the-spot approximation is used which assumes that all recombination radiation is absorbed locally.  The radiation is emitted from all star particles and is coupled to the gas via photoionisation, photoheating, and direct UV radiation pressure.  Since the radiation is advected using an explicit solver, the simulation time step is limited by the RT-Courant condition.  We have opted to subcycle the radiation time step up to 500 times per hydrodynamic time step on a given level.  In order to make this compatible with the adaptive time stepping routine used for the AMR grid, we have adopted Dirichlet boundary conditions at the interfaces between different refinement levels \citep{Commercon2014}.  Finally, we have employed the variable speed of light approximation (VSLA) presented in \cite{Katz2017,Katz2018}, that varies the speed of light used in the simulation as a function of the level in the AMR hierarchy so that the evolution of ionisation fronts is properly captured in both low and high density regimes.  The radiative transfer is only turned on when the first star forms in the simulation.

\begin{figure*}
\centerline{\includegraphics[scale=0.335]{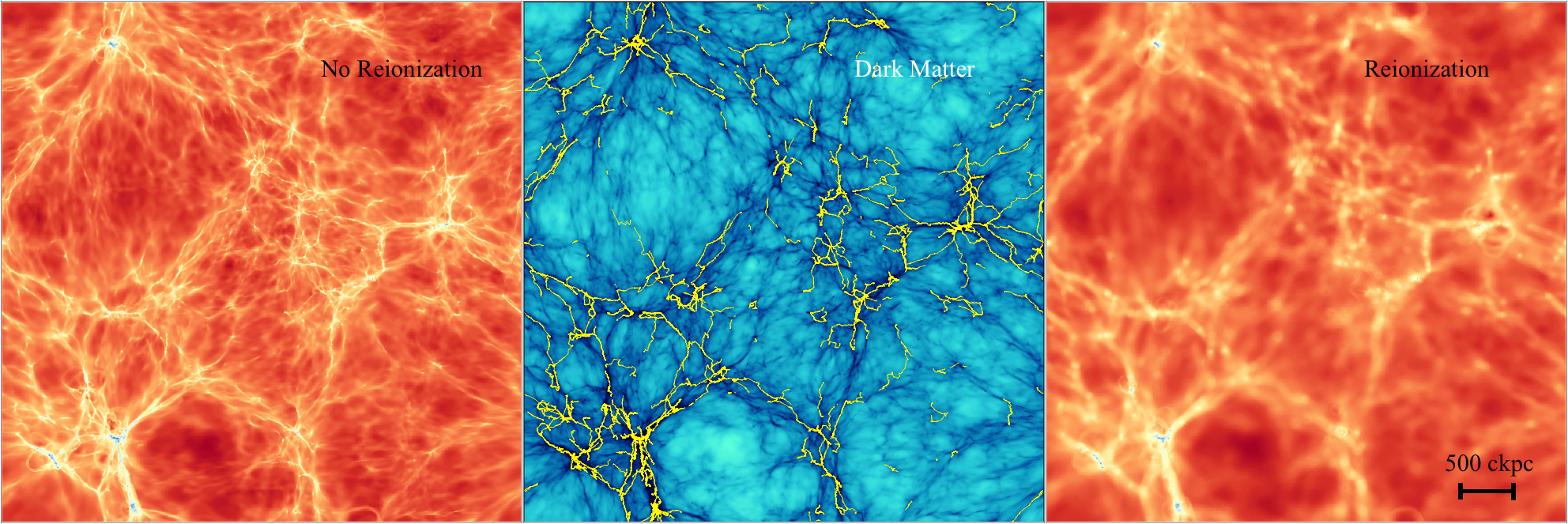}}
\caption{ {\bf (Left)} Gas column density map in a slice through the centre of the box at $z=6$ for the simulation without stellar radiation.  {\bf (Centre)} The column density of dark matter in the same slice at $z=6$.  The filamentary skeleton extracted from this slice is plotted on top of the density field for filaments that have a persistence $>6\sigma$.   The gas filamentary structure in the left panel mimics the dark matter density field.  {\bf (Right)} Gas column density map in the same slice at $z=6$ for the simulation that includes stellar radiation.  The gas filaments are significantly wider and much of the more diffuse structure has disappeared.  In all cases, the thickness of the slice is 500~ckpc.}
\label{fillimage}
\end{figure*}

Star formation is modelled using turbulence criteria inspired by \cite{Federrath2012}. Stars can only form in a cell when the gas motion is locally convergent, the cell is a local density maximum, the turbulent Jeans length is less than one cell width, and the density of the cell is greater than 200 times the cosmological mean density.  Star particles, each representing a stellar population with a minimum mass of $1,000{\rm M_{\odot}}$, are then created with a variable efficiency per free-fall time by sampling from a Poisson distribution as presented in \cite{Kimm2017,Trebitsch2017}. In this work, we only study the simulations that model stellar radiation using a spectral energy distribution (SED) consistent with binary stellar populations ({\small BPASS}) \citep{Eldridge2008,Stanway2016}.  The reionization history for this simulation and the evolution of the photoionization rate is shown in Figure~\ref{QHI}.  The {\small SPHINX} suite also contains a series of simulations that have been run with an SED that does not include binary stars; however, these simulations do not reionize the volume by $z=6$ and are thus not appropriate for studying the impact of the UV radiation on the formation of dwarf galaxies and filaments.  For comparison, we will use an identical simulation to the one described that does not include the radiation from stars so that we can isolate the effects of radiation feedback.  Note that we have saved $\sim3$ times fewer outputs for this simulation compared to the one that includes stellar radiation.

As star particles age, we impart mass, momentum, and metals into the nearby gas to model type~II supernova (SN) explosions.  This is done over the first 50Myr of the lifetime of a star particle by sampling a delay-time distribution for a Kroupa initial mass function \citep{Kroupa}.  Depending on the density, temperature, ionisation fraction, and metallicity of the gas, a variable amount momentum is injected into the nearby cells determined by the phase of the SN that is locally resolved.  Details of the algorithm can be found in \cite{Kimm2015,Kimm2017,Rosdahl2018}.  

\subsection{Halo Finding}
\label{halofinding}
Halos are identified in the simulation snapshots using the {\small ADAPTAHOP} algorithm \citep{Aubert2004,Tweed2009} in the most massive sub-maxima (MSM) mode. We define our haloes as the region inside a sphere where the mean density is 200$\rho_{\rm crit}$ centred on the densest region of the halo.  For this work, we only consider haloes with more than 300 particles (i.e. $M_{\rm vir}>6.24\times10^6$M$_{\odot}h^{-1}$).  Furthermore, to isolate the effects of the radiation feedback from all others, we will limit our analysis to a subset of isolated, main haloes.  We consider a halo to be isolated when it resides at least six virial radii away from the edge of a more massive halo \citep{Okamoto2008}.  For each halo, we assign a stellar and baryonic mass and this is computed directly from the simulation outputs as the star particles and gas cells that are within $R_{200}$ of the centre of the halo (the centre is defined as the point of maximum density).  Gas inflow and outflow rates are computed for each halo using the gas cells that have distances from the centre of $\pm5\%R_{\rm vir}$.

\subsection{Filament Extraction}
Filaments are extracted using the open-source {\small DIScrete PERsistent Structures Extractor (DisPerSE)} \citep{Sousbie2011,Sousbie2011b,Sousbie2013}.  Dark matter particles for both the simulation with and without reionization are interpolated onto a mesh using a 3D Delaunay tessellation \citep{Schaap2000} at four different redshifts ($z=9,8,7,$ and 6).  The {\small mse} routine is then used to calculate the Morse-smale complex of the dark matter density field.  Briefly, the routine identifies all critical points (i.e., points where the gradient of the density field is zero) in the 3D density cube.  This includes density maxima (type 3), filament-type saddle points (type 2), and minima (type 0).  It furthermore identifies the integral lines that connect critical points and are tangent to the gradient at all points.  The filamentary structures in the simulation are traced by the ascending 1-manifolds.  Each simulation contains millions of individual filament segments.  Each filament segment is assigned a persistence determined by the ratio of the density at the maxima and at the saddle. In general, filament segments with higher persistence connect to higher mass galaxies.  We remove low persistence pairs (below $5\sigma$) which we consider to be noise and filament segments are smoothed by averaging their positions over 30 nearest neighbours.  We also do not consider filament segments close to the edge of the box in order to remove any edge effects in the calculation of the filament properties.  Furthermore, we only study filament segments that are not within the virial radius of a halo.  During our analysis, we always bin filament segments by persistence level and only use bins that have more than 20,000 segments to avoid any biases due to low number statistics.  This selection criterion also tends to remove the highest $\sigma$ persistence filaments for which we lack statistics.  These filaments would be least impacted by reionization.

In order to assign gas properties to each filament segment, we first project the gas onto a uniform 1024$^3$ cube.  For each filament segment we rotate the box such that the axis along the filament is perpendicular to the orientation of the box and re-centre the box to the highest density gas cell within a small radius.  This re-centring is performed because sometimes the maximum density of the gas filament is offset from the maximum density of the dark matter.  We then measure radial profiles of each filament segment by averaging different gas quantities in cylindrical shells up to a distance of 90 gas cells away from the centre of the filament.  A more detailed description of our filament extraction procedure will be described in Ramsoy~et~al.~2019~({\it in prep.}).

\begin{figure*}
\centerline{\includegraphics[scale=1]{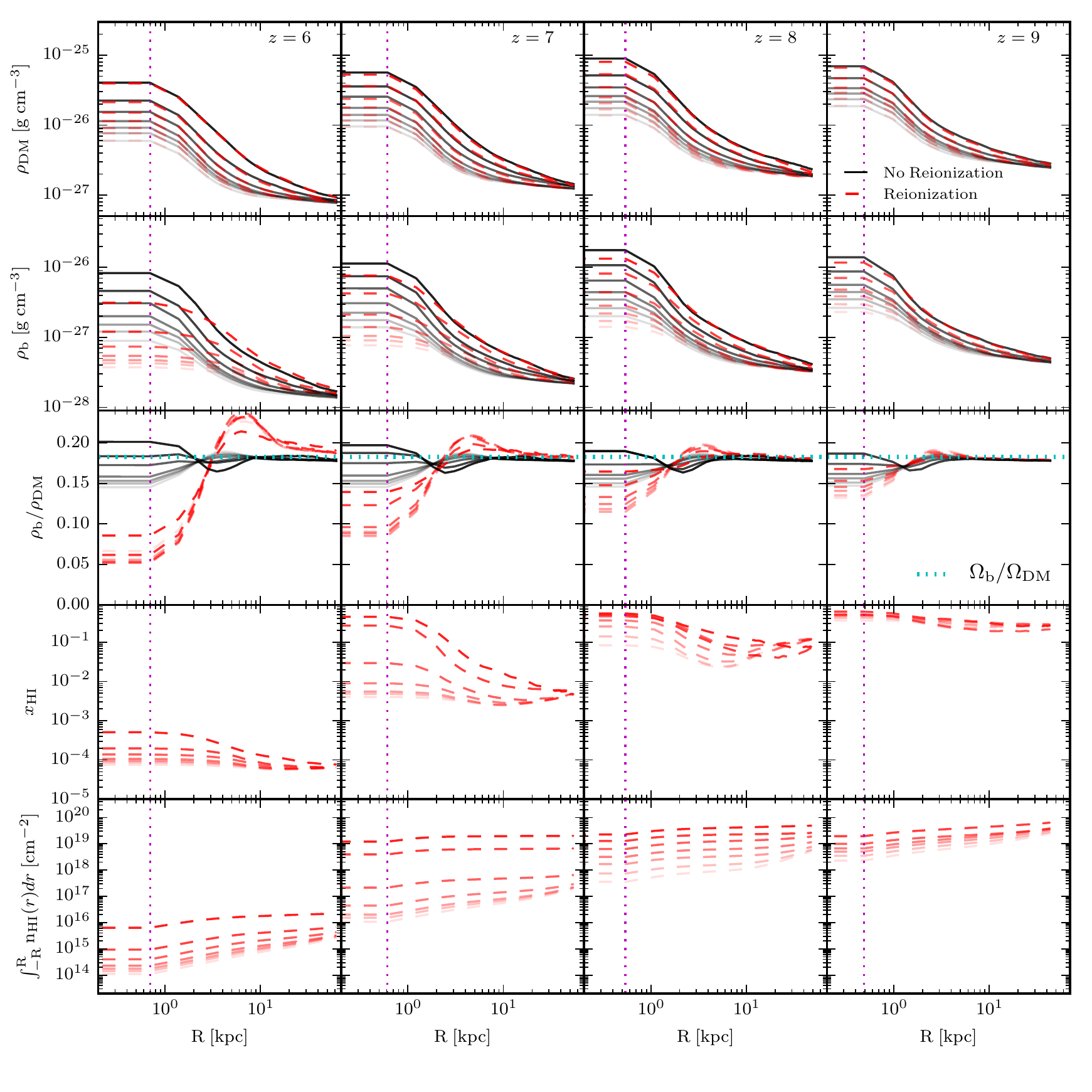}}
\caption{Dark matter density profiles (first row), gas density profiles (second row), ratio of baryonic density to dark matter density (third row), neutral fraction (fourth row), and HI column density (fifth row) of filaments in the simulations with and without stellar radiation at different redshifts.  The radius represents the cylindrical radius, perpendicular to the filament.  The red and black lines represent the simulations with and without stellar radiation, respectively, and thus with and without reionization.  The multiple lines in each panel represent filamentary structures with different persistences, with lower persistences indicated by more transparent lines.  The lowest set of lines represents structures with persistence $>5\sigma$ and the persistence increases by $0.5\sigma$ for each subsequent set of lines.  In order to dampen noise, a persistence level is only shown if it contains at least 20,000 segments.  The vertical dotted lines indicate the resolution limit of the filament extraction.  At all redshifts, the dark matter filamentary structure agrees well between the simulations with and without reionization at all persistences.  This demonstrates that reionization has a negligible effect on the large scale structure of dark matter.  In contrast, there are strong differences when gas properties are compared. }
\label{DMfil}
\end{figure*}

\section{Results}
\label{results}
In this Section, we discuss how the process of reionization impacts both the properties of intergalactic filaments and galaxies.  

\subsection{Filament Properties}
At high redshift, galaxies obtain most of their gas through accretion along filaments \citep[e.g.][]{Keres2005} and thus, in order to understand how the process of reionization impacts galaxy formation, it is important to understand how radiative feedback impacts cosmic filaments.  In the centre panel of Figure~\ref{fillimage}, we show a map of dark matter density at $z=6$ across the entire simulation box.  The left and right panels show the gas density in the simulations without and with stellar radiation, respectively.  A subset of the extracted skeleton (persistence $>6\sigma$) is traced on the dark matter map in yellow.  The locations of the extracted filaments visually trace the most prominent topological features.  We have placed a significance cut of $5\sigma$ in our analysis in order to avoid probing noise and this threshold only neglects the most diffuse filaments.  Note that in 2D, it is often difficult to visually distinguish true filaments from walls and thus certain features that may appear as filaments in 2D are indeed projected walls and thus not selected by {\small DisPerSE}.

The gas distribution in the left panel of Figure~\ref{fillimage} nearly exactly mimics the dark matter distribution.  Without photoionisation heating, the intergalactic baryons remain extremely cold, cooling adiabatically as $(1+z)^2$ upon temperature decoupling from the CMB radiation.  With little pressure support, the gas motions are dominated by the gravitational potential of the dark matter and thus there is little bias between the distribution of intergalactic dark matter and intergalactic gas without a large scale heating process.  In contrast, when stellar radiation permeates throughout the simulation box and reionizes and photoheats the IGM, pressure effects become important.  The intergalactic gas filaments in the right panel of Figure~\ref{fillimage} are significantly more diffuse and extended compared to those in the centre panel.  Furthermore, the least prominent gas filaments associated with the least robust dark matter filaments have completely disappeared in the simulation that includes stellar radiation.  This large scale smoothing of the density field is well established in the context of uniform UV backgrounds (see e.g. \citealt{Pawlik2009}) and characterising this smoothing observationally may give insight into the thermal history of the IGM \citep[e.g.][]{Miralda1994,Hui1997,Rorai2013}.

While the process of reionization strongly impacts the gas distribution, we find that radiation feedback has no impact on the large-scale distribution of dark matter.  In the top row of Figure~\ref{DMfil}, we compare the median density profiles (i.e., median density in each radial bin) of dark matter filaments of different persistence (binned in $\Delta\sigma=0.5$) between $6\leq z\leq9$ in the runs with and without reionization.  The properties of the dark matter are very similar between the two simulations, both at the onset of reionization ($z=9$) and $\sim100$Myr after reionization has completed ($z=6$).  The dark matter density profiles exhibit power-law behaviour as a function of radius \citep[e.g.][]{Aragon2010} and the lower persistence filaments, that in general feed the lower mass galaxies, have both a lower central density and smaller radius compared to the higher persistence filaments.  Note that the densities of the central regions of the filaments are likely suppressed in our simulation due to finite spatial resolution.  The resolution at which we extract the filaments is shown as the dotted vertical lines in Figure~\ref{DMfil}.  Since filaments are bound in two dimensions but not in the third, their central densities decrease as a function of redshift as the Universe expands. 

\begin{figure*}
\centerline{\includegraphics[scale=1]{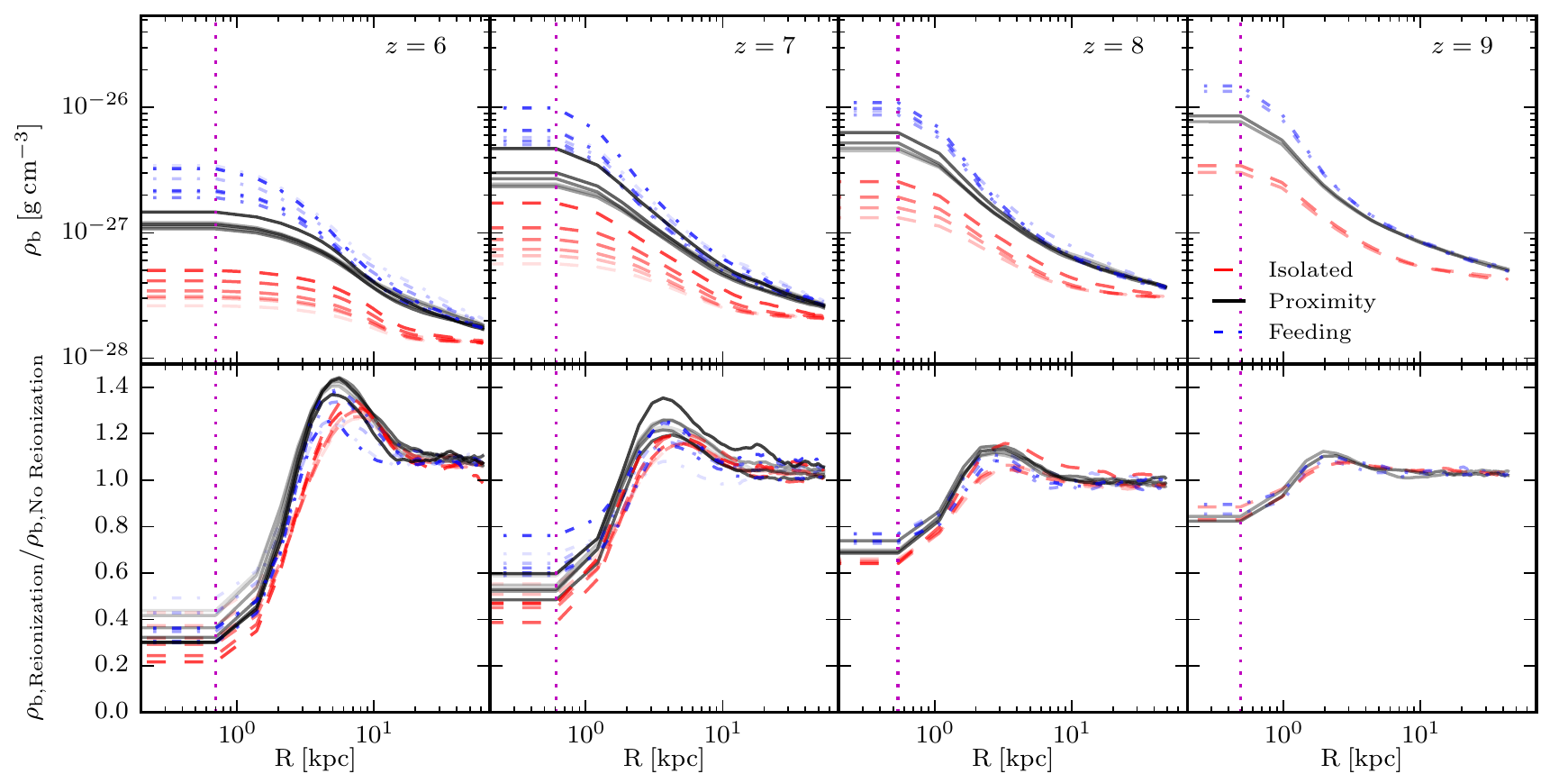}}
\caption{{\bf (Top row)} Radial profiles of gas density for filament segments in the simulation with reionization at different redshifts, split based on their distance to a halo.  Isolated, proximity, and feeding filaments (defined in the main text) are shown in red, black, and blue, respectively.  {\bf (Bottom row)} Radial profiles of the ratio of gas density in the simulation with reionization to the gas density in the simulation without reionization at different redshifts, again split into the three categories of filaments as in the top row.  The vertical dotted lines indicate the resolution limit of the filament extraction.  By $z=6$, the median central gas densities in the simulation with reionization can be reduced by 80\% compared to the simulation without radiation feedback.}
\label{GODMfil}
\end{figure*}

Turning to the gas properties of the filaments, in the second row of Figure~\ref{DMfil}, we compare the median gas density profiles between $6\leq z\leq9$ for various persistence levels in the simulations with and without reionization.  At $z=9$, when the IGM is mostly neutral, the median gas properties of the filaments are very similar between the two simulations with and without reionization.  As reionization progresses, the IGM is heated and the additional pressure support both prevents gas from accreting onto the filament and forces gas out of existing filaments.  While the gas density profiles in the simulation without reionization exhibit a similar power law profile as the dark matter, by $z=6$, the central density profiles of the gas filaments in the simulation that includes reionization have considerably flattened due to the extra pressure support.  At larger distances from the filaments, the density remains higher in the simulation with reionization compared to the simulation without as gravity still causes the gas to move towards the filament but the extra pressure prevents it from condensing.  This is exemplified in the third row of Figure~\ref{DMfil} where we show the median ratio between baryonic density and dark matter density as a function of radius for filaments of different persistence between $6\leq z\leq9$.  At $z=9$, during the onset of reionization, a small suppression in the ratio of $\rho_{\rm b}/\rho_{\rm DM}$ is already seen in the central regions of the filaments in the simulation that includes reionization compared to the simulation without.  As redshift decreases, this suppression becomes more significant and the central ratio of $\rho_{\rm b}/\rho_{\rm DM}$ can be reduced to less than $1/3$ of the cosmological value.  At larger radii ($\sim5$kpc), there is a strong enhancement in $\rho_{\rm b}/\rho_{\rm DM}$ in the simulations with radiation as gravity pulls the gas towards the filament but the additional pressure support from photo-heating prevents confinement in the dark matter filament.  In the simulation without reionization, we see some scatter in the central ratio of $\rho_{\rm b}/\rho_{\rm DM}$ due to a combination of resolution and the efficiency of cooling.  At very large radii from the centres of the filaments, boths sets of simulations converge back to the expected cosmological baryon fraction.

Since the process of reionization is inhomogeneous, it is important to understand how, in particular, filaments that are feeding galaxies are affected by reionization.  For this reason, we have split the total population of filament segments into three categories:
\begin{enumerate}
\item {\bf Feeding Filaments}:  Filament segments that are within $1-3$ virial radii of a halo.
\item {\bf Proximity Filaments}:  Filament segments that are within $3-6$ virial radii of a halo.
\item {\bf Isolated Filaments}:  Filament segments that are $>6$ virial radii away from the closest halo.
\end{enumerate}
Note that in order to classify each filament segment, we have calculated the shortest distance from each filament segment to a halo rather than distance along the filament.  

In the top row of Figure~\ref{GODMfil}, we show the median gas density profiles for the three sets of filaments, split between different persistence levels and by redshift.  In general, we find that the filaments that are actively feeding galaxies have the highest central densities, followed by the proximity filaments, while the isolated filaments tend to be the least dense.  Once again, as these systems are not gravitationally bound, their central density decreases with the expansion of the Universe.  

As reionization progresses in the simulation, ionisation fronts expand away from galaxies and into the low-density regions, before finally ionising the filaments \citep[e.g.][]{Finlator2009}.  This is easily observed in the fourth row of Figure~\ref{DMfil} which shows that the filaments become more ionised as redshift decreases.  Close to the galaxies, the photoionisation rate is much higher than in the IGM and thus one may expect that the filament segments that are close to the galaxies, are most affected from radiation feedback compared the isolated filament segments.  However, the central densities of the isolated filaments are about an order or magnitude lower than the feeding filaments and since the recombination rate scales as density squared, such filaments are considerably less self-shielded.  The balance between the photoionisation rate amplification near sources and the decreasing filament central densities away from sources determines which class of filaments is most affected by the radiation feedback.  We can measure this empirically and in the bottom row of Figure~\ref{GODMfil}, we show the ratio of the median gas density profiles with and without reionization for the three sets of filaments, split between different persistence levels and by redshift.  At $z=9$, when the simulation is only a few percent ionised, we see no difference between the different classes of filaments and we find that the global reduction in central density (or mass) is 20\%.  As reionization progresses, the different classes of filaments evolve differently and a clear separation appears at $z=7$.  At this redshift, the isolated filaments are more impacted by the radiation feedback and their central densities are reduced by 60\%.  In contrast, the denser, feeding filaments, are less impacted and a reduction of 40\% is seen.  This effect continues and by $z=6$, in all classes of filaments, the central density (and mass) of the filaments has been reduced by 60\%-80\% with respect to the simulation without reionization.  The relative difference in the reduction of central density between the different classes of filaments decreases as a more uniform radiation background permeates the IGM.

\begin{figure*}
\centerline{\includegraphics[scale=1,trim={0 0 0 0.8cm},clip]{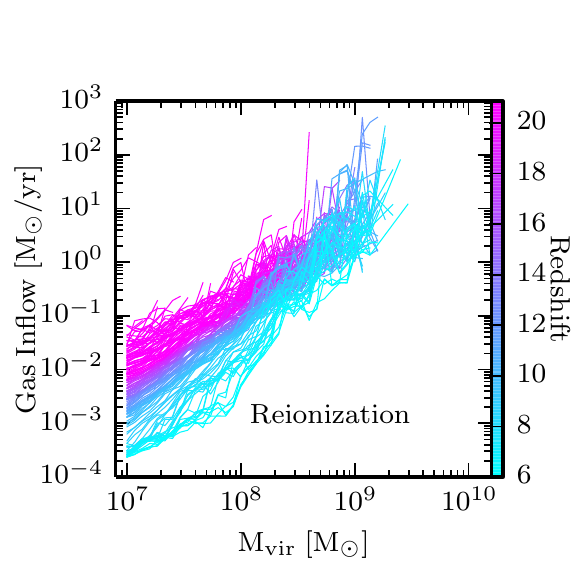}\includegraphics[scale=1,trim={0 0 0 0.8cm},clip]{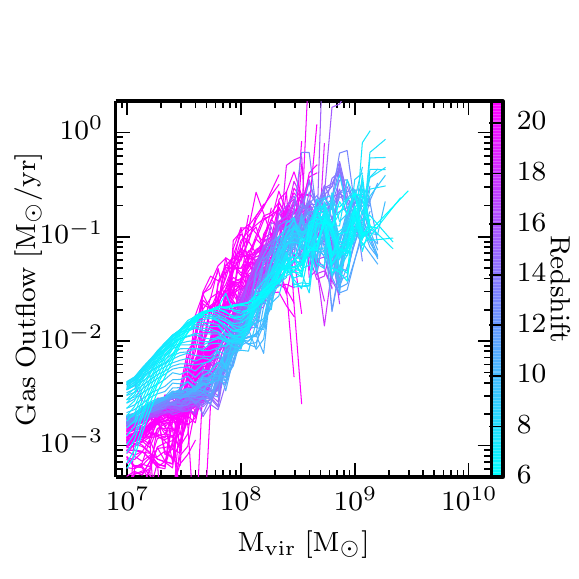}\includegraphics[scale=1,trim={0 0 0 0.8cm},clip]{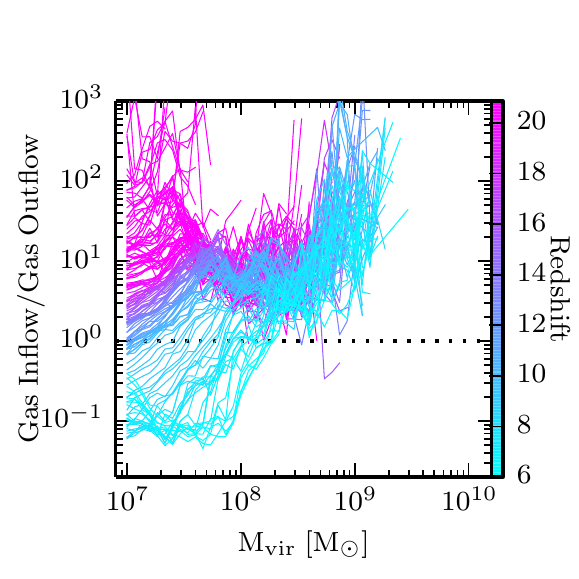}}
\centerline{\includegraphics[scale=1,trim={0 0 0 0.8cm},clip]{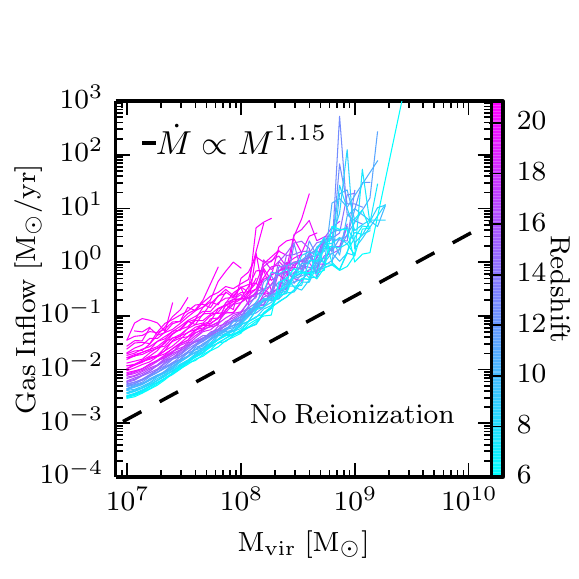}\includegraphics[scale=1,trim={0 0 0 0.8cm},clip]{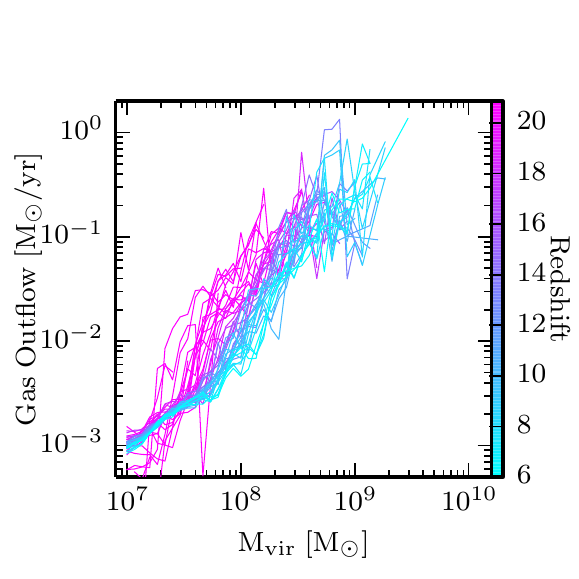}\includegraphics[scale=1,trim={0 0 0 0.8cm},clip]{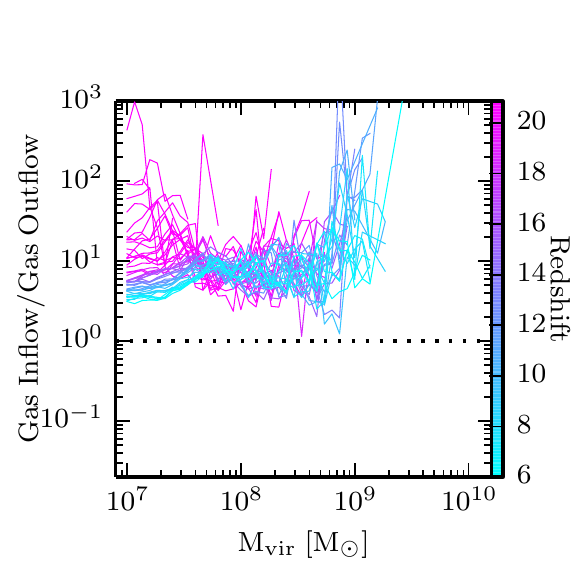}}
\caption{{\bf (Left)} The running median gas inflow rates for isolated haloes as a function of virial mass, in bins of $\Delta\log_{10}({\rm M_{vir}/M_{\odot}})=0.2$.  A bin is only shown if it contains more than 10 haloes.  The different colour lines represent snapshots at various redshifts as shown in the colour bar.  We have removed systems that are undergoing mergers in order to isolate the gas accretion from filaments.  The top and bottom panels represent the simulations with and without reionization, respectively.  {\bf (Centre)} The running median gas outflow rates for isolated haloes as a function of virial mass.  {\bf (Right)} The ratio of median gas inflow rate to median gas outflow rate as a function of virial mass.  The dotted horizontal lines indicate a ratio of one where the galaxy baryonic mass is expected to remain constant.  Systems that have a ratio below unity are losing gas due to either SN feedback or radiation feedback (in the case of the simulation that includes on-the-fly RT).}
\label{infout}
\end{figure*}

Clearly, the reduction in filament central density in the simulation with radiation compared to the simulation without is extremely significant as on average, up to 80\% of the central mass is evacuated.  This will decrease the mass available for accretion onto galaxies, suppressing their growth.  Furthermore, the filaments have clearly grown wider during the process and this is observed as the enhancement in density at larger radii from the centres of the filaments in the bottom row of Figure~\ref{GODMfil}.  Narrower, denser filaments can penetrate deep into the centres of haloes before being disrupted and this is the most efficient form of accretion.  The widening of the filaments due to reionization particularly impacts the lower mass galaxies that have virial radii similar to the width of the gas filaments.  When the filament width is comparable to the size of the halo, the gas is expected to shock long before reaching the central galaxy and the accretion switches from a more efficient ``cold mode'' to a less efficient ``hot mode'' which will reduce the amount of gas inflow into the galaxy.

\subsection{Halo Properties}
In this sub-section, we analyse how reionization affects the global properties of galaxies at different masses and redshifts as well as how individual haloes respond to the reionization of their local environment.

\subsubsection{Global Properties}
\paragraph{Inflow and Outflow Rates}
Since galaxies are primarily fed by filaments at high redshift, we expect that the reduction in total gas mass in the filaments will directly translate into the gaseous properties of high-redshift galaxies.  For each halo in the simulation, we have measured the inflow and outflow rates of gas at the virial radius in each snapshot (snapshots are spaced by $\Delta t\sim10{\rm Myr}$).  In the top left panel of Figure~\ref{infout}, we show the median gas inflow rates for isolated haloes (see Section~\ref{halofinding}) as a function of virial mass, in bins of $\Delta\log_{10}({\rm M_{vir}/M_{\odot}})=0.2$, from redshifts 6-20 for the simulation with reionization.  At fixed mass, the inflow rates tend to decrease as a function of redshift.  For example, a typical system with ${\rm M_{vir}\sim10^7M_{\odot}}$ has an inflow rate of $\sim7\times10^{-2}{\rm M_{\odot}yr^{-1}}$ at $z\sim20$ and this decreases to $\sim2\times10^{-4}{\rm M_{\odot}yr^{-1}}$ by $z=6$ for the same mass system.  The lower mass systems tend to have a larger relative reduction in their inflow rates compared to the higher mass systems as a function of redshift.  The absolute gas inflow rates increase with halo mass; however, the rate at which the inflow rates increase with halo mass is steeper for systems with ${\rm M_{vir}\gtrsim10^8M_{\odot}}$.  This indicates that the inflow rates of lower mass systems are most affected by reionization, consistent with the lowest persistence filaments being more susceptible to radiation feedback.

This is more easily seen by comparing with the inflow rates in the simulation without reionization (shown in the bottom left panel of Figure~\ref{infout}).  In general, there is a tendency for the inflow rates to decrease with time in this simulation as well.  Note how the inflow rates scale with virial mass, $\dot{M}\propto M^{1.15}(1+z)^{2.25}$ \citep{Neistein2006,Dekel2009}.  Inflow rates for ${\rm M_{vir}\sim10^7M_{\odot}}$ haloes decrease by nearly an order of magnitude from redshifts 20-6 as expected.  However, this pales in comparison to the 2.5 orders of magnitude of inflow reduction we observe for the same mass range in the simulation with reionization.  In contrast, if we compare haloes with ${\rm M_{vir}\sim10^9M_{\odot}}$, there is little difference between the simulations with and without reionization, indicating that these systems are relatively immune to reionization feedback.

\begin{figure*}
\centerline{\includegraphics[scale=1]{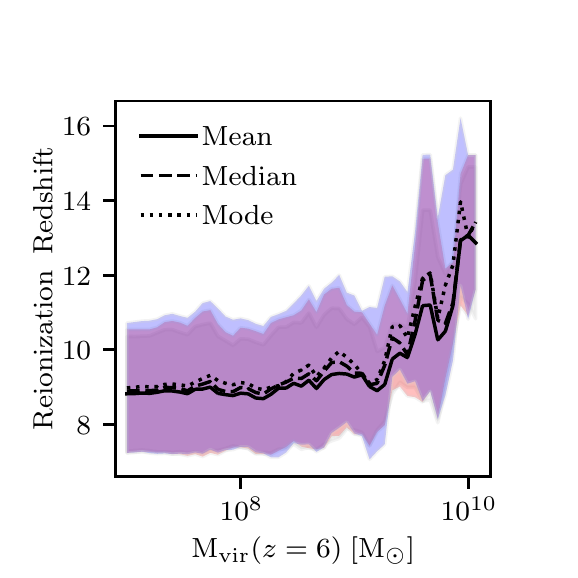}\includegraphics[scale=1]{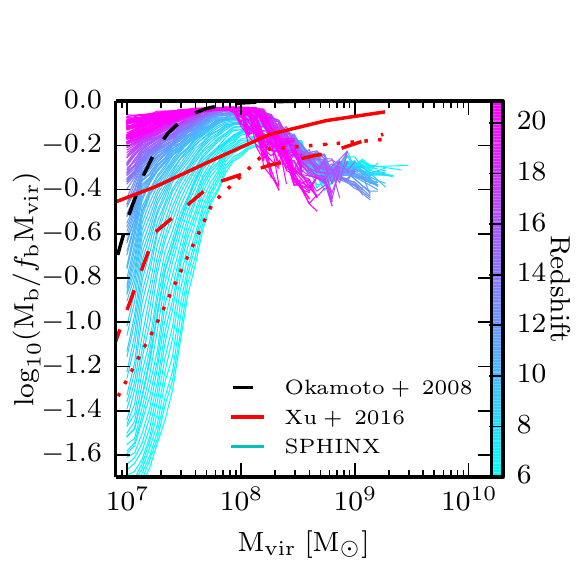}\includegraphics[scale=1]{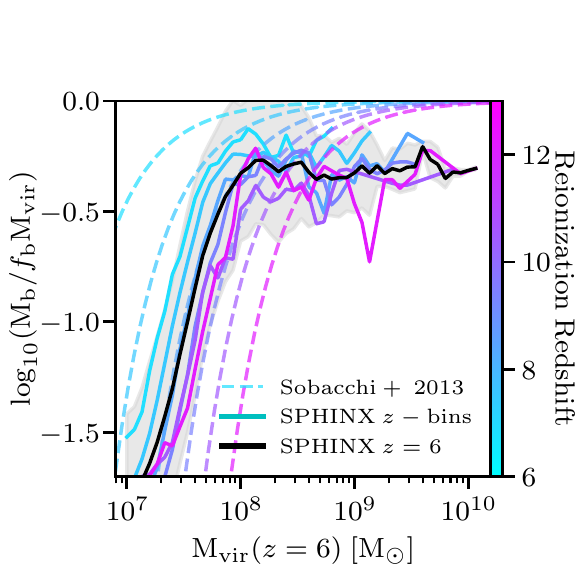}}
\caption{{\bf (Left)} The environment reionization redshift as a function of virial mass for all isolated galaxies in the simulation.  The solid, dashed, and dotted lines use the mean, median, or mode of the 27 neighbouring cells to compute the reionization redshift, respectively.  A cell is considered reionized when it reaches an ionisation fraction of 90\%.  The $1\sigma$ scatter about this relation is shown in grey, red, and blue for the mean, median, and mode, respectively.  {\bf (Centre)} Moving average in bins of $\Delta\log_{10}({\rm M_{vir}/M_{\odot}})=0.2$ of the ratio of total baryonic mass to the expected baryonic mass if the halo were to accrete its cosmic baryon fraction of gas as a function of the virial mass of the system.  A bin is only shown if it contains more than 10 haloes.  The different colour lines represent the moving average as a function of redshift and show various snapshots at redshifts in the range $6\leq z\leq 20$.  As redshift increases, the haloes host more baryons at masses $<10^8$M$_{\odot}$.  At higher masses, the ratio decreases and plateaus at a value of $50\%$ where supernova feedback regulates this fraction.  The dashed black line shows the predictions from the simulations of \protect\cite{Okamoto2008}  while the solid, dashed, and dotted red lines represent the results from the Renaissance simulations in the Rarepeak, Normal, and Void regions at $z=15$, 12.5, and 8, respectively \protect\citep{Xu2016}. {\bf (Right)}  The solid black line represents the moving average of the ratio of the total baryonic mass to the expected baryonic mass if the halo were to accrete its cosmic baryon fraction of gas as a function of the virial mass of the system at $z=6$.  The grey shaded region represents the $1\sigma$ scatter about the relation.  The solid coloured lines show the moving average of the ratio of the total baryonic mass to the expected baryonic mass for haloes at $z=6$ separated into populations where the environment was reionized at different times.  The dashed coloured lines show the predictions for the baryon fractions of inhomogeneously reionized haloes from \protect\cite{Sobacchi2013}.  The blue lines represent environments that were reionized late while the purple line represents environments that were reionized early. The haloes that reside in environments that reionized early have lower baryon fractions at a given halo mass compared to those systems which live in environments that reionized late.  All of the scatter at the low mass end can be attributed to the timing of reionization in the local environment of the halo.}
\label{bmass}
\end{figure*}

The top centre panel of Figure~\ref{infout} shows the gas outflow rates as a function of virial mass for various redshifts for the simulation with reionization.  In general, the lowest mass systems with ${\rm M_{vir}\sim10^7M_{\odot}}$ have their outflow rates increasing as redshift decreases and the simulation volume becomes more ionised.  The lowest mass systems in our simulations do not form stars and thus these outflow rates are entirely due to the galaxies being ionised from the outside-in causing the gas to escape the galaxy.  This is not the case in the simulation without reionization (see the bottom middle panel of Figure~\ref{infout}) where no amplification in outflow rate is observed for the lowest mass systems at lower redshifts.  The behaviour of the outflow rates becomes significantly more complicated at higher masses.  Systems with ${\rm M_{vir}\sim10^8M_{\odot}}$ have their outflow rates decrease between $20\gtrsim z\gtrsim13$, before increasing again down to $z=6$.  As we will discuss later, a system of this mass is likely to form stars, but as we have seen from the inflow rates, it is also susceptible to radiation feedback from reionization.  At fixed mass, galaxies that form earlier are more likely to form stars compared to those that form at later redshifts.  However, the process of reionization only occurs in our simulation at $z\lesssim11$ which may explain this unique trend in outflow rates at this particular mass.

The top right panel of Figure~\ref{infout} shows the ratio of the median gas inflow rate to outflow rate as a function of virial mass for each snapshot in our simulation with reionization.  The lowest mass systems have their inflow rates decreasing and outflow rates increasing as a function of decreasing redshift.  The ratio of these two quantities changes by four orders of magnitude between $20\gtrsim z \ge 6$.  The dotted line indicates a ratio of 1 where the inflow rates exactly balance the outflow rates.  Systems with a ratio below this value would lose their gas as a function of time.  At $z=6$, haloes with ${\rm M_{vir}\lesssim2\times10^8M_{\odot}}$ typically have a ratio $\leq1$ and are thus losing gas.  The ratio of gas inflow to outflow is approximately constant between $10^7<{\rm M_{vir}/M_{\odot}}<10^8$, with the gas outflow rate being $\sim 20$ times greater than the inflow rate.  At higher masses, this ratio increases rapidly. Even though the baryonic mass is suppressed in high mass haloes as well (see the center panel of Figure~\ref{bmass}), the inflow rate still dominates over the outflow rate for these haloes. This indicates that although the total baryon mass of the galaxy is increasing (even in the simulation without reionization), the net accumulation of DM mass is greater than that of baryons.  The outflow rates between the simulations with and without reionization are more or less consistent at a halo mass of $10^9{\rm M_{\odot}}$ and thus the heated baryons must be escaping the galaxies since we measure the outflow rates at the virial radii. Furthermore, for the haloes in the reionization simulation that have an inflow/outflow ratio below unity,  the ratio drops below unity earlier in the lowest mass systems compared to the higher mass objects.  The volume averaged photoionisation rate in the simulation increases between $6\le z\le8$ (see Figure~\ref{QHI}) and weaker UV backgrounds can only affect lower mass objects which may explain this trend.  In the bottom right panel of Figure~\ref{infout}, we show the ratio of gas inflow to gas outflow for the simulation without reionization.  For all galaxy masses, gas inflows completely dominate the system indicating that, without reionization, no system should be quenched at these high redshifts.

\paragraph{Baryonic Masses}
The reduction of gas inflow rates and increase in outflow rates due to radiation feedback significantly affects the baryonic content of galaxies during the reionization epoch.  In the centre panel of Figure~\ref{bmass}, we plot the ratio of the total baryonic mass to the expected baryonic mass (i.e. $f_{\rm b}{\rm M_{vir}}$ where $f_{\rm b}=\Omega_{\rm b}/\Omega_{\rm m}$) if the galaxy accreted its cosmic baryon fraction of gas as a function of virial mass at various redshifts.  At the highest redshifts, the galaxies exhibit baryon fractions that are very close to the expected value.  There is a slight mass dependence where the lowest mass systems tend to have lower baryon fractions.  This is partly a result of the lowest mass systems being the least resolved in the simulation and is consistent with what was found in \cite{Okamoto2008} for simulations run with a homogeneous UV background.  At the highest redshifts, the baryon fractions peak at a halo mass between $10^7<{\rm M_{vir}/M_{\odot}}<10^8$, before decreasing again.  At halo masses above this value, stars begin to form and drive outflows which decrease the total baryon content.  The turnover in baryon fraction occurs at higher masses as redshift decreases indicating that a fixed mass galaxy is less likely to form stars as redshift decreases (see Section~\ref{stellarmasses}).  Note that the exact halo mass at which this turnover occurs may be sensitive to the star particle mass in the simulation as well as the star formation and SN feedback prescriptions.  At high masses, the baryon fractions saturate at $\sim50\%$ of the cosmic value, consistent with the high-resolution radiation hydrodynamics simulations of \cite{Pawlik2017} (see also \citealt{Mitchell2018}).

The baryon fractions of the lowest mass systems change drastically between $20\gtrsim z\ge6$.  Our simulations predict that by $z=6$, a $10^7{\rm M_{\odot}}$ halo will host only $\sim2\%$ of the total expected mass in baryons.  The simulation is completely reionized at $z\sim7$ so the duration over which the baryonic fractions are significantly reduced is $\lesssim170$Myr, consistent with others who have also found this process to be rapid \citep[e.g.][]{Gnedin2014}.  The black dashed line in the centre panel of Figure~\ref{bmass} shows the expected baryon fractions as a function of virial mass just after reionization for the simulations from \cite{Okamoto2008}.  The simulations from \cite{Okamoto2008} are different from the ones in this work as they employ a homogeneous UV background and neglect the effects of self-shielding. The effect of radiation feedback on the baryon fractions of dwarf galaxies tends to be stronger in our simulations compared to what was found by \cite{Okamoto2008}, possibly due to the fact that they use a homogeneous UV background.  This is consistent with the results from the cosmological radiation hydrodynamics simulations presented in \cite{Sullivan2018} who also find a suppressed baryon fraction prior full-box reionization.  Because our simulation reionizes inhomogeneously, certain systems will have their baryonic mass depleted earlier than others, which lowers the mean baryon fraction before reionization is completed.  

The red lines in the centre panel of Figure~\ref{bmass} show the mean baryon fractions as a function of virial mass from the Renaissance simulations in the Rarepeak, Normal, and Void regions \citep{Xu2016}.  These results were measured at $z=15$, $z=12.5$, and $z=8$, respectively.  At virial masses $\lesssim10^8$M$_{\odot}$ we find fairly good agreement between the results from the Renaissance simulations and {\small SPHINX}, indicating that the impact of radiation feedback is similar between the two simulations, despite the different spatial and mass resolutions, and the different methods for modelling RT (adaptive ray-tracing versus M1).  At higher masses, we find differences between their work and ours which is likely due to different models of SN feedback.

In order to measure the effect of the inhomogeneity of reionization on the baryonic properties of high-redshift dwarf galaxies, we have measured a reionization redshift for each galaxy.  This is done by projecting the ionisation fractions onto a $64^3$ grid in a volume weighted fashion (corresponding to a cell length of $\Delta x\sim10$kpc at $z=6$, set to be larger than a typical galaxy at this redshift) and identifying the redshift at which each sub-volume is 90\% ionised.  We then use the 27 nearest neighbour sub-volumes to determine the reionization redshift of the environment around each galaxy in the simulation.  The left panel of Figure~\ref{bmass} shows the environment reionization redshift as a function of $z=6$ virial mass computed using either the mean, median, or mode of the reionization redshifts of each of the 27 neighbouring cells along with the $1\sigma$ scatter about the relations.  Any of these different methods produce reasonably consistent results in both the trend and the scatter.  We will adopt the median value for the remainder of this work.  In general, the highest mass systems at $z=6$ have their environments reionized first while the environments of the lowest mass systems reionize much later.  The typical redshift of reionization of a halo with ${\rm M_{vir}<10^9M_{\odot}}$ is $z\sim9$.  This is clearly much higher than the global reionization redshift of the simulation box at $z\sim7$. If we compare our result at $z=9$ to the prediction from \cite{Okamoto2008}, we find a much better agreement at the low mass end (where SN feedback is not dominant).  Thus the discrepancy between our result at $z=7$ and \cite{Okamoto2008} can likely be explained by the inhomogeneity of reionization in our simulation.  The scatter in reionization redshift of the environments for the systems with ${\rm M_{vir}<10^9M_{\odot}}$ is relatively constant, representing a $\Delta z\sim3.5$ which corresponds to $\Delta t\sim320$Myr (see also \citealt{Aubert2018}).  This time period is 40\% of the age of the Universe by $z=7$, indicating that the timing of reionization is highly variable from galaxy to galaxy.

\begin{figure*}
\centerline{\includegraphics[scale=1,trim={0 0 0 0.8cm},clip]{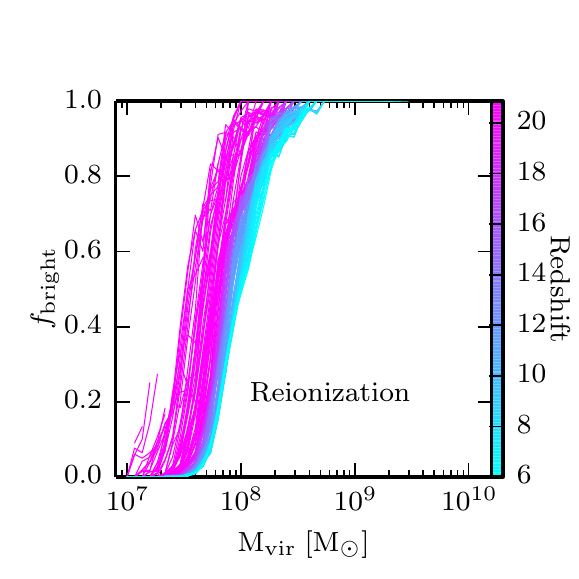}\includegraphics[scale=1,trim={0 0 0 0.8cm},clip]{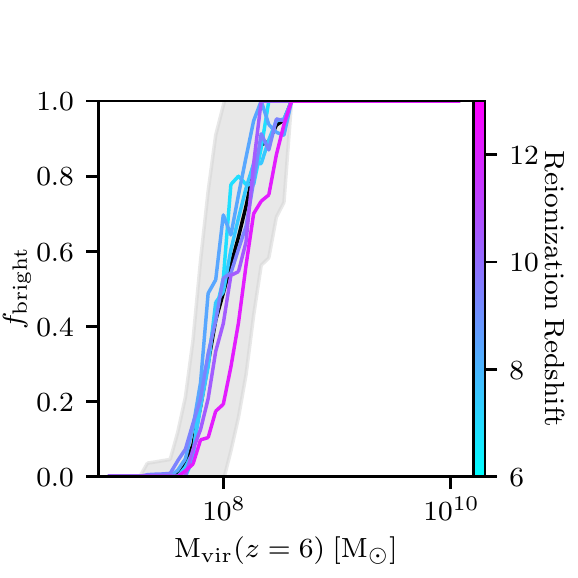}\includegraphics[scale=1,trim={0 0 0 0.8cm},clip]{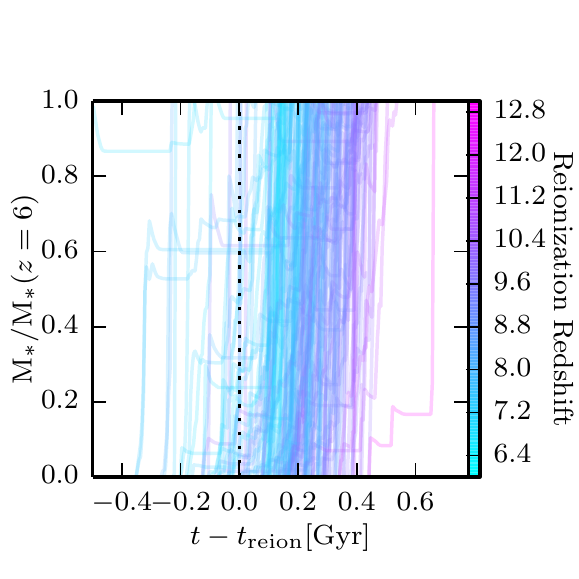}}
\centerline{\includegraphics[scale=1,trim={0 0 0 0.8cm},clip]{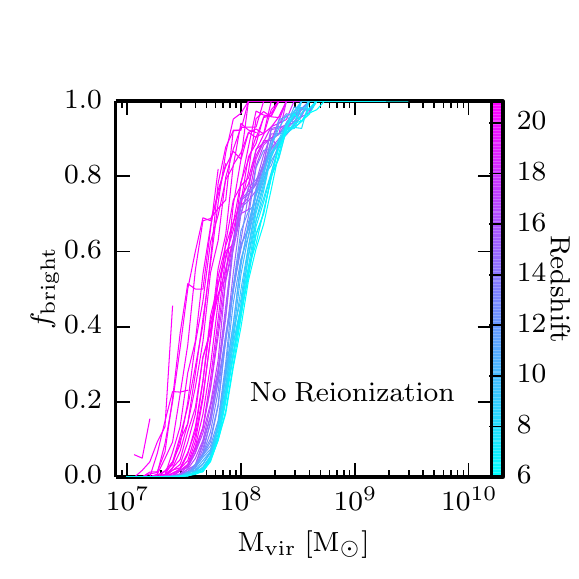}\includegraphics[scale=1,trim={0 0 0 0.8cm},clip]{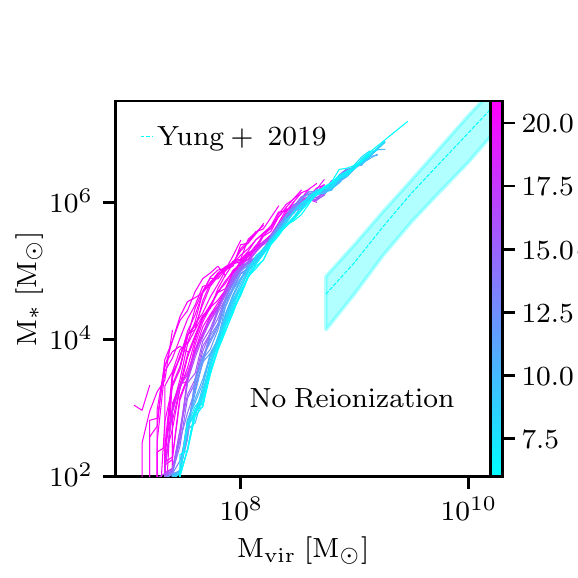}\includegraphics[scale=1,trim={0 0 0 0.8cm},clip]{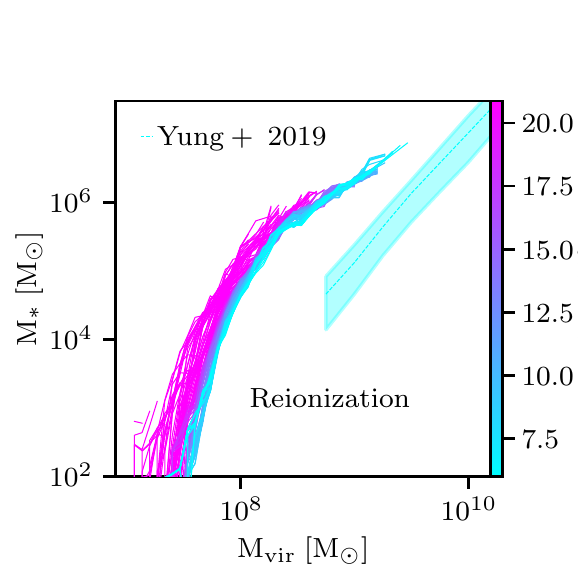}}
\caption{{\bf (Top Left)} Moving average in bins of $\Delta\log_{10}({\rm M_{vir}/M_{\odot}})=0.2$ of the fraction of halos that host at least $1000{\rm M_{\odot}}$ in stars (i.e., one star particle) as a function of the virial mass of the system for the simulation that includes reionization.  A bin is only shown if it contains more than 10 haloes.  The different colour lines represent the moving average as a function of redshift and show various snapshots at redshifts in the range $6\leq z\leq 20$.  As redshift decreases, the probability that a fixed mass system is luminous also decreases.  {\bf (Top Centre)} The solid black line represents the moving average of the fraction of haloes that host at least one star as a function of the virial mass of the system at $z=6$.   The grey shaded region represents the $1\sigma$ scatter about the relation.  The coloured lines show the moving average separated into populations where the environment was reionized at different times.  The blue lines represent environments that were reionized late while the purple line represents environments that were reionized early.  Systems that reionized earlier have a slightly lower probability of hosting at least one star particle.  {\bf (Top Right)} Fraction of $z=6$ stellar mass formed as a function of time since the environment around the halo was reionized.  The vertical dotted line indicates the time at which the environment of an individual halo was reionized.  The different colours represent haloes that reionized at different times.  We only show systems with ${M_*(z=6)/M_{\odot}<10^5}$ (which corresponds to a halo mass ${\rm M_{vir}\lesssim5\times10^8M_{\odot}}$) to select the haloes that are most likely to be affected by reionization. {\bf (Bottom Left)} Same as top left but for the simulation without reionization.  {\bf (Bottom Centre)} Mean stellar mass-halo mass relation at different redshifts for the simulation without reionization. The shaded blue region represents the $z=6$ semi-analytic model predictions from \protect\cite{Yung2019}.  {\bf (Bottom Right)} Same as bottom centre but for the simulation with reionization.}
\label{smass}
\end{figure*}

In the right panel of Figure~\ref{bmass}, we plot the baryon fraction of all isolated systems at $z=6$ as a function of virial mass along with the $1\sigma$ scatter (shown as the black line and the grey shaded region).  The scatter in baryon fraction at fixed mass is indeed large at all masses and for a system with ${\rm M_{vir}=10^8M_{\odot}}$, we would predict that the baryon fraction may fall anywhere between 20\%-100\%.  The different colour solid lines in this figure represent subsets of the galaxy population that have been split based on when their environment was reionized.  The systems that reionized the latest (shown in light blue) exhibit the highest baryon fractions at fixed mass while the systems that reionized earlier (shown in purple) scatter well below the median relation.  Interestingly, for systems with ${\rm M_{vir}\lesssim10^8M_{\odot}}$, the entirety of the scatter about this relation is driven by the reionization redshift of the environment around a galaxy.  At higher masses, radiation feedback is much less efficient at removing baryons from galaxies or preventing inflows and SN feedback dominates the outflow rates, hence the scatter is no longer dominated by the reionization redshift.  Nevertheless, for the lowest mass systems, the $z=6$ baryon fractions are an absolute probe of the reionization redshift of the environment around the galaxy. 

Even within a fixed mass and reionization redshift, there is still scatter in the baryon fraction.  Some of this is likely due to the fact that depending on environment, each halo will experience a different photoionisation rate (compared to the mean) as a function of redshift.  Other effects such as the virialization redshift of the halo, the speed of the ionisation front, as well as the orientation of the halo and filaments with respect to the ionisation front are likely to impact the baryon fraction. However, from the results of the right panel of Figure~\ref{bmass}, it seems that the reionisation redshift is the dominant effect.

The coloured dashed lines in the right panel of Figure~\ref{bmass} show the predictions of baryon fraction for haloes that reionize at different times from \cite{Sobacchi2013}.  Note that \cite{Sobacchi2013} use a spherical 1-D collapse model to explore the impact of radiation on dwarf galaxies, neglecting both self-shielding and the evolution of the UV background.  Qualitatively, their predictions are in fairly good agreement with what is found empirically in the simulations.  These equations slightly over-predict the effect for the haloes that reionize first and there is an under-prediction of the effect for the haloes to reionize last; however, for intermediate redshift of reionization, these formulae work well.  Part of the differences could be due to the fact that we have calibrated their relations to the mean photoionisation rates in our simulations when actually this value is likely higher around galaxies than in the IGM.  Furthermore, the simulations of \cite{Sobacchi2013} are 1D and do not account for the effects of reionization on the filaments which may explain why their model over-predicts the baryon fractions for the low mass haloes that reionized the latest.

In order to quantify the effects of reionization on the baryonic content of galaxies during the epoch of reionization, we provide tables for the baryon fraction as a function of virial mass and redshift in Appendix~\ref{fitfunc}.  Note that there is a critical mass above which SN feedback dominates over radiation feedback in regulating the baryonic content of galaxies.  We also provide fitting functions in Appendix~\ref{fitfunc} for this transition mass and find that at $z=7$, this threshold mass is $\sim10^{7.8}{\rm M_{\odot}}$.

\paragraph{Stellar Masses}
\label{stellarmasses}
The impact of reionization on the gas content of galaxies directly translates into their stellar properties.  In the top left panel of Figure~\ref{smass}, we show the fraction of isolated haloes that host at least $1000{\rm M_{\odot}}$ in stars ($f_{\rm bright}$), representing one star particle\footnote{We define a galaxy as having at least one star particle as being ``luminous''.}, as a function of virial mass at various redshifts.  With time, the probability that a halo at fixed mass will host at least $1000{\rm M_{\odot}}$ in stars also decreases.  This is consistent with other work that use either a homogeneous or inhomogeneous UV background \citep[e.g.][]{Sawala2016,Ocvirk2016}.  In the redshift interval $20\ge z \ge 6$, the halo mass at which the probability of being luminous is 50\% increases from $\sim3\times10^7{\rm M_{\odot}}$ to $\sim10^8{\rm M_{\odot}}$.  By $z=6$, more than 150Myr after reionization, we find that all systems with ${\rm M_{vir}\gtrsim5\times10^8M_{\odot}}$ will host a luminous galaxy (see also \citealt{Wheeler2018}).  This is a significantly smaller virial mass than what was found by \cite{Ocvirk2016} who predict that ${\rm M_{vir}}$ must be greater than $\sim2\times10^9{\rm M_{\odot}}$ at the end of reionization in order to host a luminous galaxy.  This holds even when accounting for the differences in star particle mass.  Our value is more comparable to \cite{Sawala2016}.  The major differences between the simulations that will affect haloes of this mass are spatial and mass resolution and our spatial resolution is more similar to that in \cite{Sawala2016} than that in \cite{Ocvirk2016} which may explain this discrepancy.  In other words, higher resolution tends to permit star formation in lower-mass halos even with radiation feedback.

We emphasise that radiation feedback and the process of reionization is not responsible for the evolution in $f_{\rm bright}$ towards higher halo masses in our simulations.  This is likely due to an evolution in gas density and decreasing spatial resolution.  In the bottom left panel of Figure~\ref{smass} we show $f_{\rm bright}$ as a function of virial mass and redshift for the simulation without reionization and the behaviour is nearly identical to that in the simulation that includes radiation feedback.  Comparing the stellar mass-halo mass relations between the two simulations with and without reionization (see the bottom centre and bottom right panels of Figure~\ref{smass}), we find that this relation is similar between the two simulations.  The only major difference occurs in haloes with ${\rm M_{vir}\lesssim2\times10^7M_{\odot}}$ where there are slightly more luminous haloes in the simulation without radiation feedback.  It is only in this extremely low mass regime where reionization is impacting the stellar masses of our objects. While there are no observational constrains on the stellar mass-halo mass relation at the virial masses probed in our work, it is important to note that our quantitative predictions for this relations do not agree with all others in the literature.  A more detailed analysis of this is discussed in \cite{Rosdahl2018}.  In the bottom centre and bottom right panels of Figure~\ref{smass} we show our results compared to the predictions from the semi-analytical model of \cite{Yung2019}.  We predict significantly larger stellar masses despite the fact that in both works, the agreement with the observed UV luminosity function is good.

We should note that $f_{\rm bright}$ may be somewhat sensitive to the choice of star particle mass.  In the {\small SPHINX} suite, each star particle is formed with a mass that is an integer multiple of $1000{\rm M_{\odot}}$.  The stellar mass-halo mass relation from the higher resolution (0.7pc) simulations of \cite{Kimm2017} predict that a halo with ${\rm M_{vir}=10^7M_{\odot}}$ will host at most $1000{\rm M_{\odot}}$ in stars (which is reasonably consistent with \citealt{Xu2016}), with many systems scattering to much lower stellar masses.  The stellar mass-halo mass relation of the {\small SPHINX} simulation \citep{Rosdahl2018} is in good agreement with the Renaissance simulation \citep{Xu2016} at ${\rm M_{vir}}\lesssim10^8{\rm M_{\odot}}$ where the effect of radiation is strongest (see also Figure~13 of \citealt{Cote2018}).  Because of the finite mass resolution in the {\small SPHINX} suite, we expect some stochasticity in our parameter $f_{\rm bright}$.  However, when understood this way, we find that haloes with ${\rm M_{vir}\lesssim3\times10^7M_{\odot}}$ are unlikely to host more than 1000M$_{\odot}$ mass in stars by $z=6$ which is very consistent with much higher resolution (both spatial and mass) simulations \citep{Kimm2017}.

In the top centre panel of Figure~\ref{smass}, we show $f_{\rm bright}$ as a function of virial mass at $z=6$ along with its $1\sigma$ scatter.  Similar to the baryon fractions, we find that at fixed mass, the $1\sigma$ scatter can be extremely large.  For example, the $1\sigma$ scatter at ${\rm M_{vir}=10^8M_{\odot}}$ extends from 0\%-100\%.  The different colour lines on the plot represent subsets of the population that had their environment reionized at different redshifts.  In contrast to what we found for the baryon fractions, the scatter in $f_{\rm bright}$ cannot be completely explained by the reionization redshift of the local environment around a galaxy.  There is some evidence that systems that reionized later are slightly more likely to host at least one star particle, which may be due to the higher baryonic content that these galaxies exhibit; however, clearly other processes are more important than environment reionization redshift in determining whether or not a system at fixed mass forms stars.

The top right panel of Figure~\ref{smass} shows the fraction of stellar mass formed as a function of time since the environment around a halo was reionized ($t_{\rm reion}$), for individual isolated galaxies with ${M_*(z=6)/M_{\odot}<10^5}$.  We place this upper bound on stellar mass to select the low-mass haloes that are most likely to be affected be reionization.  This stellar mass is typically hosted by a halo of mass ${\rm M_{vir}\sim5\times10^8M_{\odot}}$ at $z=6$.  Many observational campaigns have searched for a signature of reionization by examining the star formation histories of local group dwarf galaxies to identify those that formed the majority of their stars at $z>6$ \citep[e.g.][]{Grebel2004,Monelli2010,Hidalgo2011,Weisz2014,Brown2014,Skillman2017,Bett2018}.  If the majority of the galaxies in our simulation formed their stars prior to their local reionization redshifts, the individual galaxy tracks plotted in the top right panel of Figure~\ref{smass} would approach unity on the left side of the vertical dotted line.  This is clearly not the case and instead we find that for these low-mass dwarf galaxies, most of the stellar mass is formed after the environment around the halo is reionized.  A significant fraction of the expected gas content remains present in systems with ${\rm M_{vir}\sim5\times10^8M_{\odot}}$ (see the right panel of Figure~\ref{bmass}) so it is not particularly surprising that most of the stars form after the environment is reionized.  Hence we argue that reionization quenches star formation in dwarf galaxies by gradually starving them of gas rather than by photoevaporation.  This is further discussed in the next section. 

We have identified a group of haloes where we find a long plateau in stellar mass (i.e., the star formation rate is 0) for a period of time after the local environment was reionized.  What happens in these systems is that a strong burst of star formation leads to the production of copious amounts of ionising photons.  The escape fractions in our simulations are feedback regulated (see \citealt{Wise2009,Trebitsch2017,Kimm2017,Rosdahl2018}) and thus the onset of supernova creates low opacity channels for these ionising photons to escape.  Thus the timing of the supernova often corresponds to the reionization of the local environment.  The supernova and radiative feedback drives dense gas out from the centre of the galaxy while also preventing further inflows and therefore preventing further star formation until the galaxy can resettle.  Hence, the plateaus in stellar mass that we have found for some of the galaxies in our simulation immediately after $t_{\rm reion}$ correspond to those systems that had a large burst of star formation and self-ionised their surroundings (rather than being externally ionised).  However, even these systems often form more stars at some point after this catastrophic event.

Recently, \cite{Dawoodbhoy2018} used the CoDa-I simulations \citep{Ocvirk2016} to study how star formation in dwarf galaxies was suppressed by inhomogeneous reionization.  They find that in haloes with ${\rm M_{vir}\lesssim10^9 M_{\odot}}$, the star formation rates rapidly declined following the reionization of the local region.  This result contradicts what we have found with the {\small SPHINX} simulations.  The CoDa-I simulations are at considerably lower spatial resolution compared to our work and it is unlikely that these simulations are able to resolve self-shielding in the lower mass haloes.  For this reason, when an ionisation front sweeps over the galaxy, much of the gas is photo-heated to temperatures $>2\times10^4$K which is their temperature threshold for star formation.  The suppression on star formation is less extreme in the CoDa-II simulation which removed this temperature criteria \citep{Ocvirk2018} and any differences that remain between our simulation and this work can likely still be ascribed to resolution.

In order to quantify the stellar content of galaxies during the epoch of reionization, we have created analytic fitting functions for $f_{\rm bright}$ as a function of virial mass and redshift.  We have fit $f_{\rm bright}$ using a logistic function such that
\begin{equation}
f_{\rm bright} = \frac{1}{1+e^{-k[\log_{10}({\rm M_{vir}})-{\rm M_c}(z)]}},
\end{equation}
where $k$ sets the steepness of the function and ${\rm M_c}(z)$ is the $\log_{10}$ of the critical mass where $f_{\rm bright}=50\%$ as a function of redshift, $z$.  We empirically find that ${\rm M_c}(z)$ is well modelled as being linear with redshift such that
\begin{equation}
{\rm M_c}(z) = cz+d.
\end{equation}
We have fit for $k$, $c$, and $d$ and find that $c=-0.05$, $d=8.32$, and $k=8.20$ provide a reasonably good approximation to the data from $20\geq z\geq6$.  For more exact values of $f_{\rm bright}$ as a function of virial mass and redshift, we provide a table in Appendix~\ref{fitfunc} with the exact values and $1\sigma$ scatter for this quantity.

\begin{figure*}
\centerline{\includegraphics[scale=1,trim={0 0.8cm 0 1.0cm},clip]{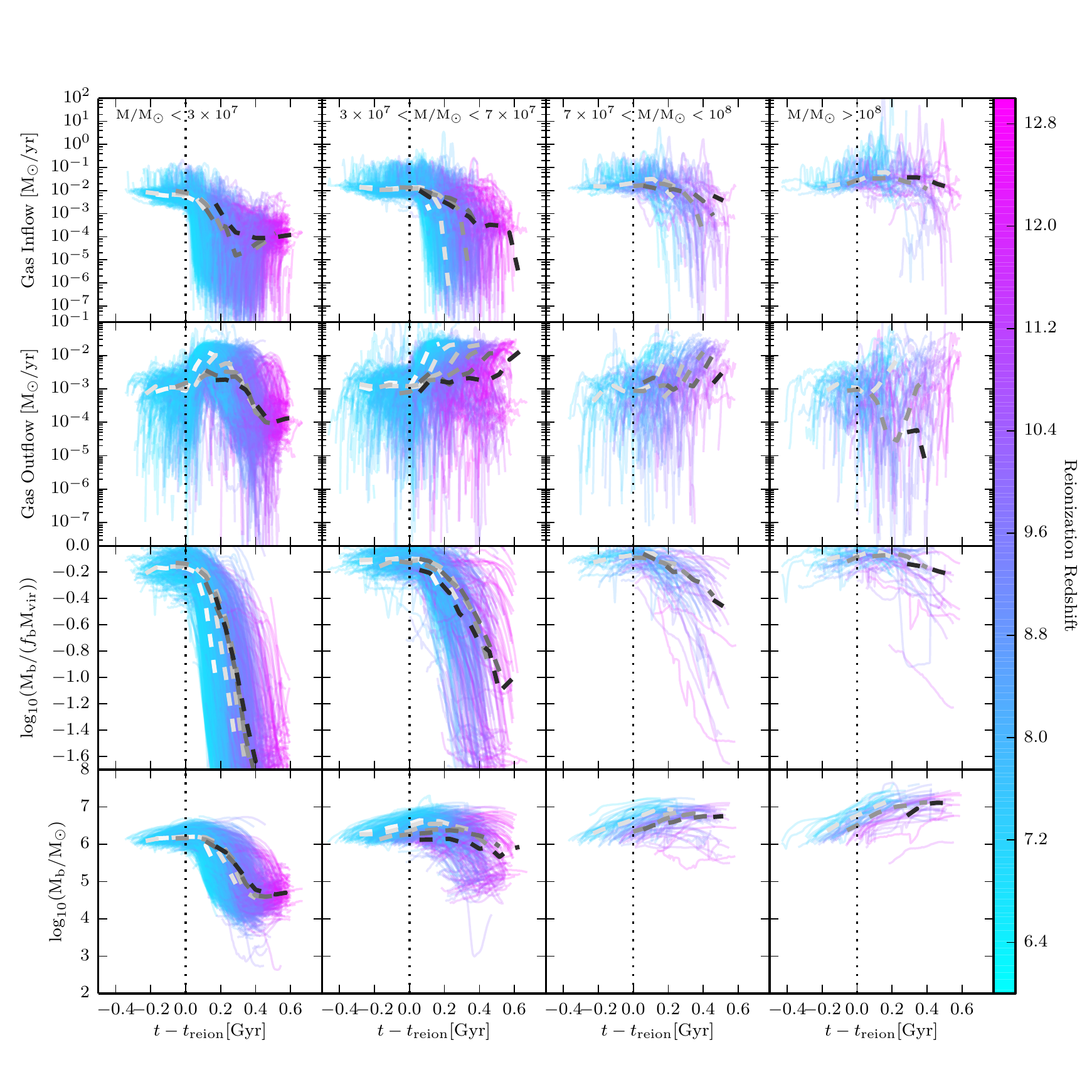}}
\caption{Gas inflow rates (first row), gas outflow rates (second row), the ratio of total baryonic mass to the expected baryonic mass if the halo were to accrete its cosmic baryon fraction of gas (third row), and total baryonic mass (fourth row) as a function of time since environmental reionization for individual galaxies selected for being both ``isolated'' and for not hosting stars.  The colour of the lines depicts the reionization redshift of the environment around the halo.  The four columns represent different mass bins as indicated on the plots.  The different dashed lines represent the median values of different subsets of galaxies at fixed mass that reionized at different times.  Shading from white to black represents systems that reionized progressively earlier (i.e., white were reionized the latest while black were reionized the earliest).  The dotted vertical lines indicate the time of reionization and the haloes are binned by their mass at $z=6$.}
\label{infsp}
\end{figure*}

\subsubsection{Individual Haloes}
In this section, we further analyse the effects of reionization on the formation of dwarf galaxies by considering only galaxies that are both ``isolated'' and do not host any stars which allows us to break any degeneracy between the effects of stellar feedback and radiation feedback.  Using the halo merger trees, we have tracked the properties of each halo in the simulation as a function of time and study in this section how the baryonic properties of this special population of galaxies responds to reionization.  We have separated the population of galaxies into four mass bins based on their mass at $z=6$: ${\rm M_{vir}/M_{\odot}}<3\times10^7$, $3\times10^7<{\rm M_{vir}/M_{\odot}}<7\times10^7$, $7\times10^7<{\rm M_{vir}/M_{\odot}}<10^8$, and ${\rm M_{vir}/M_{\odot}}>10^8$, corresponding to circular velocities of $\sim7.5,\ 10.6,\ 13.0,\ {\rm and}\ 15.9$~km/s, respectively.  The number of galaxies in each of these mass bins are 1979, 553, 84, and 65, respectively.

\paragraph{Inflow Rates}
In the top row of Figure~\ref{infsp}, we show how the gas inflow rates for individual galaxies evolve as a function of time since the environment around the halo was reionized ($t_{\rm reion}$).  The colour of the tracks indicate the redshift at which the environment around the halo reionized.  In general, the inflow rates of the lowest mass systems respond to reionization more quickly than the higher mass systems.  The dashed lines, which represent the median relation for galaxies in bins of different reionization redshift, decrease more quickly after $t_{\rm reion}$ in the first two panels of the top row of Figure~\ref{infsp} compared to the latter two.  For the lowest mass systems with ${\rm M_{vir}<3\times10^7M_{\odot}}$, it takes only 200Myr before the inflow rates have been decreased by two orders of magnitude compared to their pre-reionization values.  There is a large scatter among individual galaxies regarding how much the inflow rates are suppressed with some exhibiting a suppression of 6 orders of magnitude.  The median relation often does not decrease by such a large factor because within a given redshift bin, there is a range in suppression times, and after reionization, accretion can restart if the gas cools to low enough temperatures.

Interestingly, within a fixed mass bin, the systems that have the latest environmental reionization times tend to have their inflow rates suppressed more quickly compared to those systems that reionized earlier.  This effect can be seen most clearly in the first two panels of the top row of Figure~\ref{infsp}, where the white and light grey dashed lines, representing the late reionizing systems have a downturn that occurs closer to $t_{\rm reion}$ compared to the dark grey and black curves, representing the early reionizing systems.  This could be due to a number of factors.  The strength of the UV background tends to increase in our simulations at $z\lesssim8.5$ and thus the filaments should reionize more quickly due to the higher photoionisation rate.  Observations suggest that the photoionisation rate should continue to increase until $z\sim4$ \citep{Calverley2011,Wyithe2011}, so we expect this process to become even quicker if the simulation were continued to lower redshifts.  Furthermore, the filaments are becoming less dense as redshift decreases and thus, even for a fixed photoionisation rate, one would expect that the filaments to be more susceptible to radiation feedback at lower redshift, thereby decreasing the inflow rates onto galaxies more quickly.  

\paragraph{Outflow Rates} 
As the inflow rates are suppressed and the photoionisation rate around the halo increases, we expect that the outflow rates for individual haloes will increase in response to an increase in gas pressure \citep[e.g.][]{Rees1986,Shapiro2004,Iliev2005}.  In the second row of Figure~\ref{infsp}, we plot the outflow rates as a function of time since environmental reionization for individual galaxies selected to be both ``isolated'' and without stars.  This selection criteria ensures that the measured outflow is due to ``external'' radiative feedback rather than internal stellar feedback.  Regardless of halo mass, the median outflow rates tend to peak at $\sim2\times10^{-2}{\rm M_{\odot}yr^{-1}}$.  For the lower mass systems, this value exceeds the median inflow rates (see the top row of Figure~\ref{infsp}) and thus these galaxies are expected to lose baryonic mass, consistent with what was observed in the previous section when analysing the entire galaxy population.

Similar to the inflow rates, we find that the galaxies that have their environments reionized later tend to have their outflow rates increase much more quickly after $t_{\rm reion}$ compared to these systems that reionized earlier.  Note how the lighter lines increase more rapidly after $t_{\rm reion}$ compared to the darker lines in the centre two panels of the second row of Figure~\ref{infsp}, showing that galaxies that reionized later have more responsive outflow bursts.  This is likely to be due to two factors: firstly, as we have indicated previously, the strength of the volume weighted photoionisation rate is increasing towards the end of reionization and through the post-reionization epoch and therefore, at fixed density, the gas will be ionised more quickly.  Secondly, the virial radius is changing (due to a decrease in critical density) such that the gas density at the virial radius will decrease with redshift and thus become more susceptible to radiation feedback.

\begin{figure*}
\centerline{\includegraphics[scale=1]{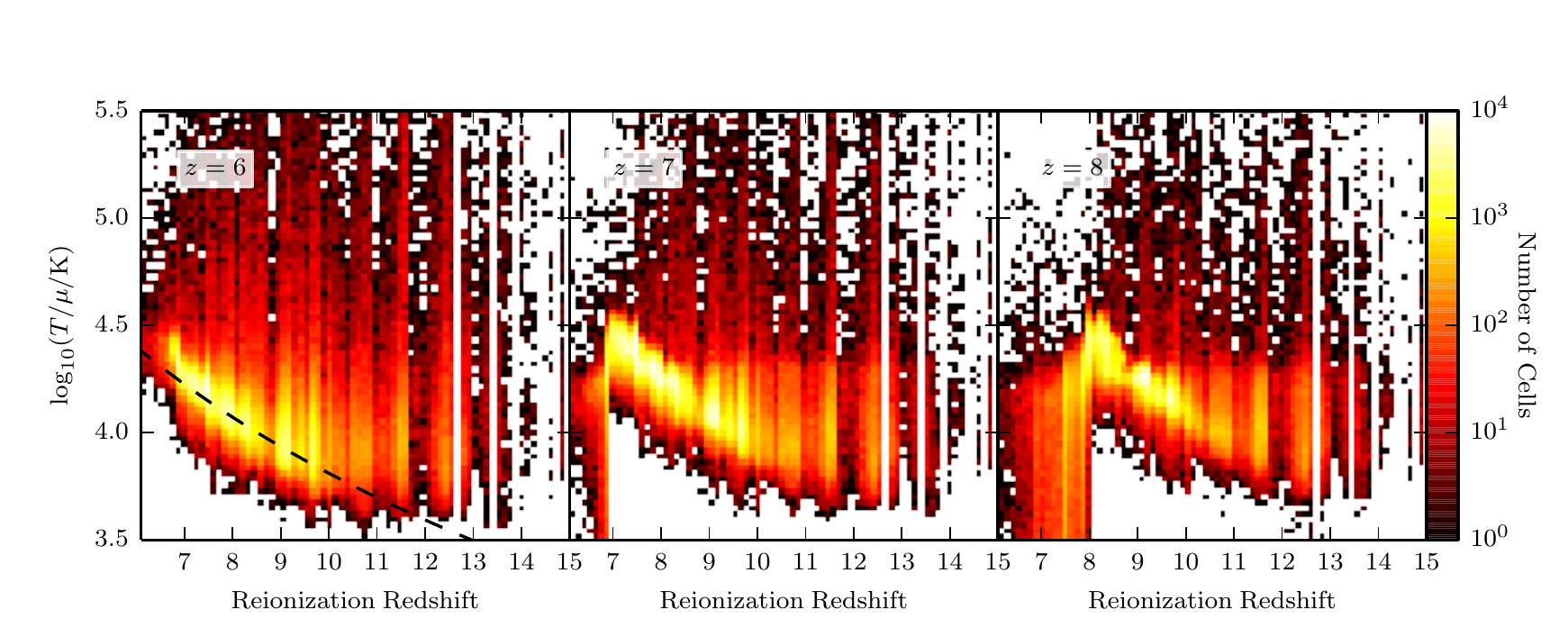}}
\caption{Gas temperature in the simulation versus reionization redshift for gas cells in the simulation.  The left, centre, and right panels show the temperature at $z=6$, $z=7$, and $z=8$, respectively.  The cells (represented by those in the coarsened 64$^3$ volume) that have been recently reionized have a temperature of $\sim25,000$K and this gas can cool to nearly 5,000K long after the cell was reionized.  The dashed black line in the left panel represents an adiabatic cooling curve for cosmic expansion.}
\label{temptot}
\end{figure*}

The shapes of the median curves in second row of Figure~\ref{infsp} provide an indication of how efficiently reionization expels mass from the halo.  For all systems with ${\rm M_{vir}\gtrsim3\times10^7M_{\odot}}$, the outflow rates at $z=6$ (indicated by the last point in each curve) remain above their pre-reionization values suggesting that the process of gas expulsion is continuing.  In contrast, we find that the lowest mass systems with ${\rm M_{vir}\lesssim3\times10^7M_{\odot}}$ have very low outflow rates long after the local environment is reionized.  The outflow rates for these systems are maximised at $\sim200$Myr after $t_{\rm reion}$ before decreasing to relatively low values.  From the right panel of Figure~\ref{bmass}, we can see that this halo mass exhibits $\lesssim3\%$ of its expected quantity of baryons.  The inflow rates have all been strongly suppressed hence there remains very little additional gas to expel from the halo.  The more massive haloes have not quite reached this regime because their inflow rates are less sensitive to reionization, and likewise, they can become more dense since they are more massive and self-shield better than lower mass haloes.  Nevertheless, the outskirts of even the more massive systems are somewhat sensitive to photoevaporation.  Comparing the magnitude of the change in outflow rates in the second row of Figure~\ref{infsp} to the change in inflow rates in the first row of Figure~\ref{infsp}, it is clear that the dominant mechanism that regulates the baryonic content of a high-redshift halo is the suppression of inflows.

\paragraph{Baryonic Masses}
The changes to the inflow and outflow rates of these ``isolated'', star-free haloes manifest themselves in the baryon fractions of the haloes.  In third row of Figure~\ref{infsp}, we plot the fraction of expected baryons (${\rm M_b/f_bM_{vir}}$) as a function of time since environmental reionization.  Due to the suppression of inflow rates and the enhancement of outflow rates after $t_{\rm reion}$, it is no surprise that we begin to see a reduction in the baryon fractions at this time.  This effect becomes weaker with increasing halo mass as identified previously.  In the right panel of Figure~\ref{bmass} we showed that the scatter in baryon fraction at fixed mass is entirely due to the timing of reionization in the local environment of the halo.  From the left two panels of the third row of Figure~\ref{infsp}, which show the baryon fractions for the lowest mass systems, we observe that at the times when the median lines truncate (indicating $z=6$), the baryon fractions are higher for the systems that reionized later.  This can be explained by the fact that the baryon fraction is a time integrated quantity and the systems that reionized earlier have their inflow rates suppressed and outflow rates enhanced for a longer period of time.  Despite the fact that these quantities respond faster to radiation feedback in the later reionizing systems, this is not enough to overcome the timing differences.  Note that for a system with ${\rm M_{\rm vir}<3\times10^7M_{\odot}}$, it takes $\sim200$Myr for the baryon fraction to decrease to 10\% of the cosmic value if the system was reionized at $z<7$; however, is that same system reionized at $z>11$, this process takes $\sim350$Myr.  

For haloes with ${\rm M_{\rm vir}>3\times10^7M_{\odot}}$, the difference in response time of the baryon fraction to environmental reionization is less pronounced compared to either the lower mass systems or the inflow/outflow rates.  This may be due to the fact that for the latest reionizing systems, the decrease in baryon fraction is a much weaker effect for these higher mass systems compared to those in the left panel of third row of Figure~\ref{infsp}, and that response time of the inflow and outflow rates to reionization takes longer for the higher mass systems.  The dominant effect which drives the scatter in the right panel of Figure~\ref{bmass} is the time since reionization rather than the response time to reionization.  If we allow the simulation to progress further and the baryon fraction to decrease for the higher mass systems, we expect that this effect will appear for at least those in the mass range $3\times10^7<{\rm M_{\rm vir}/M_{\odot}<7\times10^7}$.  The reduction in baryon fraction by $z=6$ for systems with ${\rm M_{\rm vir}\sim10^8M_{\odot}}$, is almost negligible, and peaks at $\sim40\%$ for those systems that reionized the earliest.  Most of the ``isolated'' and star-free haloes of this mass in our simulation have virial temperatures below $10^4$K indicating that the atomic cooling threshold is likely not an adequate approximation for the threshold mass to be sensitive to radiative feedback from reionization, rather it is considerably lower.  Note that the {\small SPHINX} simulations reionize slightly early compared to what is observed and the volume-weighted photoionisation rate at $z=6$ is higher in the simulation than what is observed \citep{Rosdahl2018}.  Therefore, the simulations may slightly over-predict the effect of reionization on dwarf galaxies at $z=6$.  For this reason, the true threshold halo mass that is sensitive to reionization at $z=6$ is likely less massive than even what we see in our simulations and thus has a virial temperature that is potentially well below the atomic cooling threshold.

\begin{figure*}
\centerline{\includegraphics[scale=1]{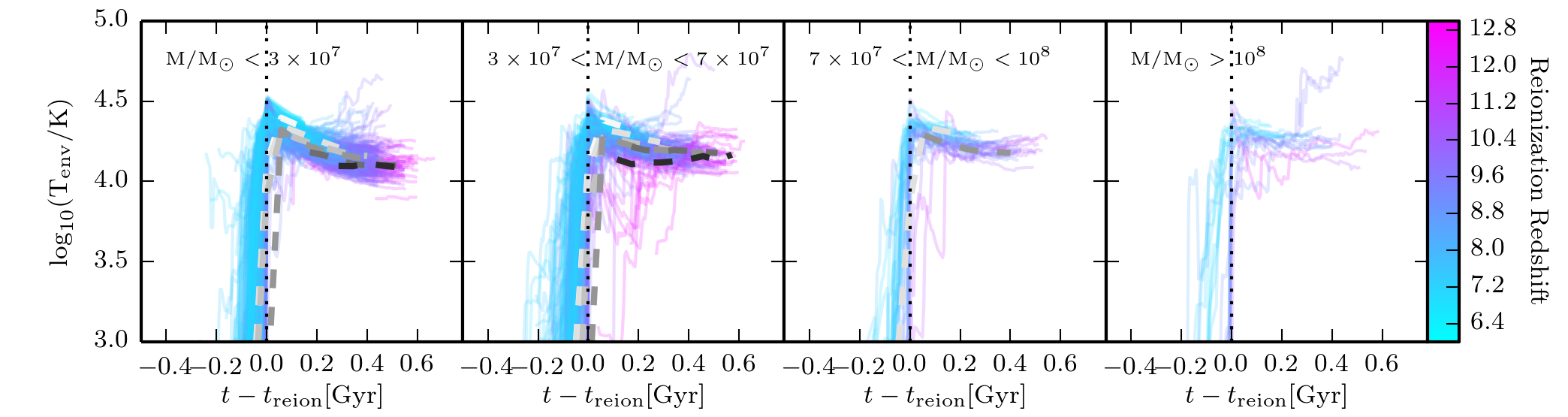}}
\caption{Gas temperature in the environment around a halo as a function of time since environmental reionization for the most ``isolated'' and star-free individual galaxies.  The colour of the lines depicts the reionization redshift of the environment around the halo.  The four panels represent different mass bins as indicated on the plots.  The different dashed lines represent the median values of different subsets of galaxies at fixed mass that reionized at different times.  Shading from white to black represents systems that reionized progressively earlier (i.e., white were reionized the latest while black were reionized the earliest).  The dotted vertical lines indicate the temperature at the time of reionization.}
\label{temphist}
\end{figure*}

Additionally, we should emphasise that there is a different characteristic mass for systems that have their accretion shut down versus those haloes which lose gas due to photoevaporation.  Nearly all haloes with ${\rm M_{vir}<10^8M_{\odot}}$ suffer some degree of inflow suppression as shown in first row of Figure~\ref{infsp}.  We argued that the reduction of the gas inflow rate was much stronger than the enhancement of the gas outflow rate and thus inflow suppression is the dominant effect of reionization compared to photoevaporation.  In the fourth row of Figure~\ref{infsp}, we show the total baryonic mass for ``isolated'', star-free haloes as a function of time since environmental reionization.  On average, for all haloes with ${\rm M_{vir}>3\times10^7M_{\odot}}$, the total baryonic mass contained within the halo in the post-reionization epoch is either greater than or equal to the total baryonic mass contained within the halo before the environment was reionized.  This indeed suggests that photoevaporation is a weak effect for haloes with ${\rm M_{vir}>3\times10^7M_{\odot}}$.  In contrast, haloes with ${\rm M_{vir}<3\times10^7M_{\odot}}$ lose a significant portion of their baryonic mass due to photoevaporation.  Prior to environmental reionization, these haloes tend to contain $\sim10^6{\rm M_{\odot}}$ in gas and this value decreases to $\lesssim10^5{\rm M_{\odot}}$ within $\sim400$Myr after environmental reionization.

\subsection{Thermodynamic State of the IGM}
Various authors have attempted to explain the critical mass at which radiation feedback inhibits gas accretion onto haloes by comparing the temperature of the gas at various over-densities to either the virial temperature of the halo, or the temperature of the gas at the virial radius \citep[e.g.][]{Gnedin2000b,Hoeft2006,Okamoto2008}.  Understanding the temperature evolution of the IGM is of key importance for determining whether or not a halo can accrete gas because, as we have shown, the properties of the filaments that feed the haloes are sensitive to the temperature evolution.  If the temperature of the surrounding gas falls below the shock temperature at the edge of the filament, then it is more likely to be captured by the filament and cool to the core, if the density is high enough (see \citealt{Ocvirk2016}).  In principle, this condensation within filaments should lead to higher accretion rates onto haloes. 

In Figure~\ref{temptot}, we plot a 2D histogram of gas temperature versus reionization redshift for all cells in the simulation at $z=6$, $z=7$, and $z=8$ \footnote{Note that we degrade the simulation (using a volume weighted scheme) to a $64^3$ grid to compute the temperature, as was done for the reionization redshift, in order to create Figure~\ref{temptot}.}.  Due to the chosen SED of sources in our simulation, upon ionisation, the gas is photoheated to $T\sim25,000$K.  Certain cells in all three panels of Figure~\ref{temptot} scatter to values much higher than this ionisation temperature because they have been affected by a recent supernova.  In the centre and right panels of Figure~\ref{temptot}, most of the gas that reionized later than $z=7$ and $z=8$, respectively exhibits a temperature $T<10^4$K as this gas has yet to be photoheated and much of it has been adiabatically cooling since the start of the simulation.  After the gas has been reionized, it will adiabatically cool due to the expansion of the Universe and it is evident in all three panels that all gas that was ionised prior to the redshift shown has cooled to $T<25,000$K.  The spatial
fluctuations of  the temperature-density relation we observe in our simulations are a direct result of the inhomogeneity of reionization \citep[e.g.][]{Daloisio2015,Keating2017}.  By $z=6$, much of the gas that was reionized by $z=10$ has cooled to a temperature $T<10^4$K.  This process is expected to continue until HeII reionization occurs, photoheating the gas again at lower redshift \citep[e.g.][]{Puchwein2018}.

\begin{figure*}
\centerline{\includegraphics[width=15.7cm]{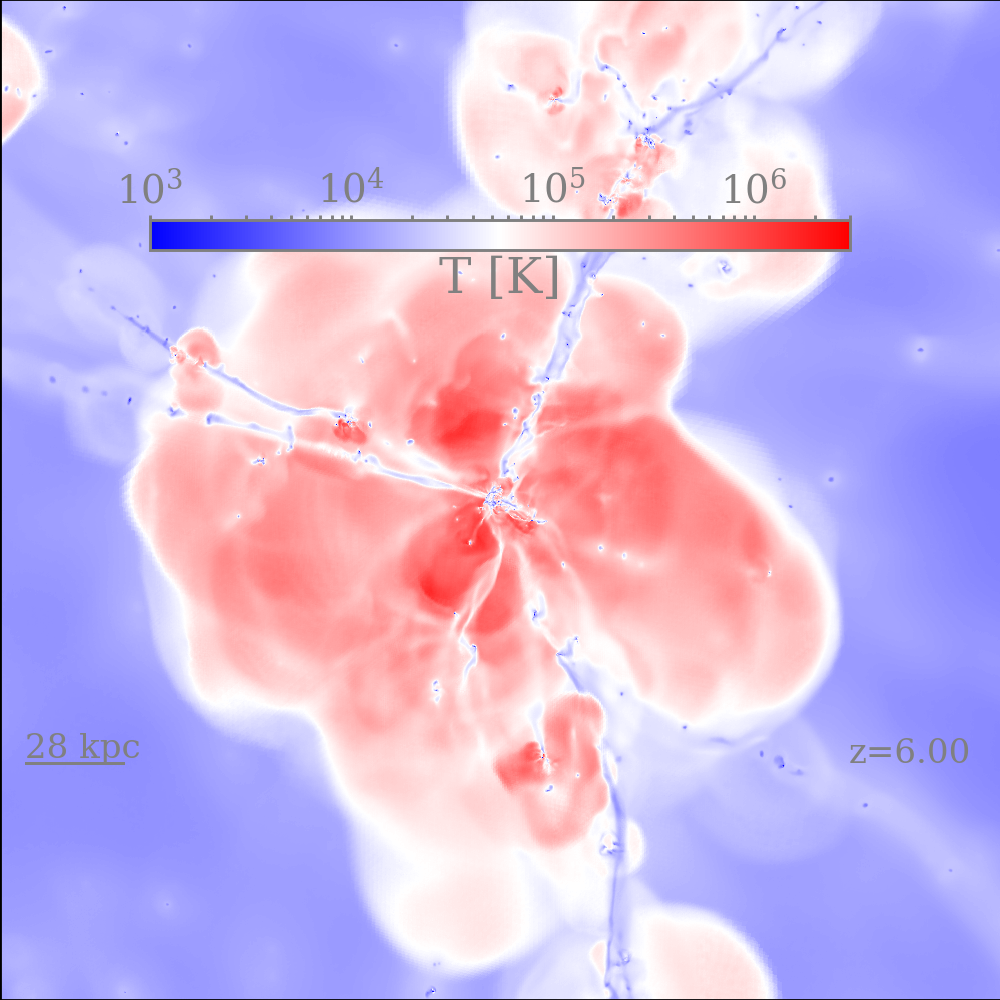}}
\caption{Gas temperature map of the region surrounding a massive galaxy at $z=6$ in our large 10Mpc volume.  The red regions represent gas heated by SN while the blue regions represent photoheated gas.  In the post-reionization epoch, the filaments feeding massive galaxies are dense enough to still have cool cores and can continue to efficiently feed gas to the centres of haloes.}
\label{awesome}
\end{figure*}

In order to understand how gas is cooling near the galaxies, we select the most isolated, starless haloes in the simulation (i.e. those that are at least 10 virial radii away from a larger halo) in order to mitigate any effects from SN on the temperature of the surrounding medium, and plot in Figure~\ref{temphist} the temperature of the environment around a halo (taken as the median temperature of the 27 closest cells to the halo in the $64^3$ grid) as a function of time since the environment around the halo was reionized.  Prior to environmental reionization, the gas temperature has a value $T\ll10^3$K; however, this increases nearly instantaneously at the time of reionization (denoted by the dotted vertical line) to $T\sim25,000$K.  Gas cooling is clearly evident after the environment has reionized for the majority of the lowest mass haloes with ${\rm M_{vir}<3\times10^7M_{\odot}}$.  The surrounding temperatures approach $T\sim10^4$K at $t-t_{\rm reion}\sim500$Myr.  This cooling signature of the gas around the halo is required for some of the systems to begin accreting gas again.  

The gas in the local environments around haloes cools less than what is seen for much of the gas in the IGM shown in Figure~\ref{temptot}. Most of the IGM can cool to $T<10^4$K.  The gas at mean density and below represents most of the volume of the Universe and is thus the dominant contribution to Figure~\ref{temptot}.  Post-reionization, this gas asymptotically approaches a power law temperature-density relation as discussed by \cite{Hui1997}.  In contrast, the gas much closer to haloes has a density that is higher than average and may require more than one photon per baryon to remain ionised throughout the first billion years.  While the strength of the UV background is high enough to maintain ionisation around these systems, additional photoheating occurs even after the cell is reionized as photons are being absorbed to maintain the ionisation state. 

From the top row of Figure~\ref{infsp} we find no evidence that the inflows for individual low-mass galaxies return to their pre-reionization levels long after reionization.  Even with the gas cooling in the IGM, the shock temperatures at the edge of the lowest persistence filaments are still well below the mean temperature and thus the gas is not efficiently recaptured.  In the simulation without reionization, we do observe such low temperature shocks.  The decrease in the central density of the filaments combined with the widening of the gas filaments due to the extra pressure support means that the lowest mass systems are unlikely to be able to ever efficiently accrete gas following the reionization of their local environment, unless they grow significantly in halo mass.  Hence we expect that the majority of these low mass galaxies will remain starved of gas for the remainder of the evolution of the Universe.  In contrast, \cite{Ocvirk2016} has shown that massive filaments which feed high mass galaxies develop strong shocks that heat the gas to temperatures above which the IGM is photoheated to during reionization and thus have cool cores (see also \citealt{Kaurov2015}).  In these filaments, the gas can cool to the core of the filament and efficiently feed a galaxy.  A similar effect is also seen in our simulations (see Figure~\ref{awesome}).  Hence any filament with a deep enough gravitational potential to compress the photoheated gas in the post-reionization epoch should remain relatively immune to the effects of radiation feedback.  

\section{Discussion \& Conclusions}
\label{discon}
In this work, we have exploited the {\small SPHINX} simulations, a suite of full-box high-resolution cosmological radiation hydrodynamics simulations, to understand how radiation feedback from the process of reionization impacts the gas content of the Universe and suppresses dwarf galaxy formation.  This topic has been widely discussed in the literature \citep[e.g.][]{Rees1986,Gnedin2000b,Shapiro2004,Iliev2005,Hoeft2006,Okamoto2008,Sobacchi2013,Gnedin2014,Noh2014,Ocvirk2016,Xu2016,Pawlik2017,Sullivan2018,Wu2019} for numerous reasons including its implications for the ``missing satellite problem'' \citep{Klypin1999,Moore1999,Bullock2000} and the ability to regulate reionization \citep[e.g.][]{Iliev2007}.  Various recent high-resolution cosmological radiation hydrodynamics simulations are now agreeing that star formation can occur in haloes with masses as low as $10^7{\rm M_{\odot}}$ \citep{Wise2014,Xu2016,Kimm2017}, well below the atomic cooling threshold, which leaves open the question of the exact impact of reionization on the gas and stellar content of the lowest mass haloes that may have similar properties to the progenitors of the locally observed ultra-faint dwarf galaxies.  Furthermore, it is only recently that the effects of inhomogeneity of reionization can be studied in a full-box cosmological radiation hydrodynamics simulation \citep[e.g.][]{Dawoodbhoy2018}.  

Here, we use the state-of-the-art {\small SPHINX} simulations to elucidate some of the physics that governs how the gas and stellar properties of high-redshift dwarf galaxies responds to radiation feedback.  We do this by first systematically studying the impact of reionization on inflows through cosmic filaments which is the primary means by which high-redshift galaxies obtain their gas.  We then analyse how the baryon fractions and stellar content of haloes respond to the rapid changes in filament properties.  Finally, we follow individual haloes throughout cosmic time to understand how the gas fraction, stellar content, and environmental temperature around a halo change due to the reionization of the local environment.  

Certain caveats should be kept in mind when interpreting our results.  In order to obtain the high spatial and mass resolution needed to resolve the lowest mass haloes in a full-box cosmological radiation hydrodynamics simulation, our simulation volume is only $5^3$~comoving Mpc$^3$ which indicates that our simulation may not be truly representative of the high-redshift galaxy population.  We have tried to mitigate this by selecting initial conditions that are as ``average'' as possible \citep{Rosdahl2018}; however, by not having a larger box, we exclude larger mass galaxies that can drive stronger UV background fluctuations than are currently captured by our simulation. Capturing all of the physics that occur on the largest scales of reionization would require boxes that are $\sim100\times$ the size of the ones used in this work.  The quantitative results, in particular the exact mass at which the turnover in baryon mass occurs due to reionization as well as the scatter in reionization times at fixed halo mass might change if we could capture the full large-scale process of reionization.  We also note that our simulation reionizes the volume earlier than observations suggest which could impact the results quantitatively.  Furthermore, haloes of ${\rm M_{vir}\sim10^7M_{\odot}}$ are only resolved by a few hundred dark matter particles and thus resolution may still play a role in suppressing the baryonic content of the lowest mass haloes in our simulation. A full convergence study will need to be completed to determine how the critical mass at which reionization impacts the baryon masses quantitatively changes at different spatial and mass resolutions; however, we note that \cite{Okamoto2008} show good convergence in this critical mass at $z=5$ using simulations that have resolutions approximately equal to and higher than the ones used in this work.  We do not include a model for Population~III star formation and have assumed that the simulation volume has been uniformly metal enriched by a generation of stars that has not been modelled\footnote{This crudely approximates the regime where H$_2$ dominates.}.  If this enrichment is patchy or to a different level than what we have assumed, gas cooling in the lowest mass haloes may change which will impact their ability to form stars.  Likewise, molecular hydrogen, which may dominate the cooling rate in these haloes at high redshift is not explicitly modelled by our simulations.  Future work will ideally consider all of these drawbacks in our current effort, however, we expect, qualitatively, that the physics we have outlined in this work will hold regardless of the box size and additional subgrid models.

With these caveats in mind, our conclusions can be summarised as follows:
\begin{itemize}

\item While dark matter filaments are unaffected by reionization, photoheating in the IGM due to reionization pressurises the gas which significantly widens and removes much of the gas content in the least robust filamentary structures.  By $z=6$, the median decrease in the gas mass at the centres of filaments due to reionization is 60-80\% depending on the proximity of the filament to a halo (isolated haloes are more affected than haloes directly feeding galaxies).  Filaments that have shock temperatures greater than the post-reionization IGM gas temperature, which tend to feed high-mass galaxies, can still maintain cool cores and efficiently feed galaxies.

\item The gas inflow rates onto haloes with ${\rm M_{vir}\lesssim10^8M_{\odot}}$ respond to the change in filament properties and are reduced by more than an order of magnitude compared to simulations without radiation feedback.  Similarly, gas outflow rates increase for the lowest mass haloes with ${\rm M_{vir}\lesssim3\times10^7M_{\odot}}$ due to photoevaporation.  Haloes that reionize later have their inflow and outflow rate respond quicker to reionization compared to haloes that reionize earlier due to a higher flux of ionising photons and decreased background density.  For almost all halo masses, the reduction of inflows is a more important process than photoevaporation and hence starvation is the dominant process by which reionization quenches star formation in dwarf galaxies.  

\item Reionization is inhomogeneous and affects the more massive systems in the Universe before the lowest mass systems.  In our simulations, we find that the average redshift of reionization for a halo of ${\rm M_{vir}=10^{10}M_{\odot}}$ is $z\sim13$ while this decreases to $z\sim9$ for a halo with mass ${\rm M_{vir}=10^{8}M_{\odot}}$.  The scatter in reionization redshift at fixed mass is significant, $\Delta z\gtrsim3$, although the exact value may be sensitive to the size of the simulation volume. 

\item Because of the inhomogeneity of reionization there is significant scatter in the baryon fraction of haloes at fixed virial mass in the post-reionization epoch.  For haloes with ${\rm M_{vir}\lesssim10^8M_{\odot}}$, the entirety of this scatter can be attributed to the inhomogeneous reionization redshifts of individual haloes. 

\item Since inflow suppression is more important than photoevaporation, haloes of all masses can form stars long after their environment has been reionized.  This implies that there should be no signature of an instantaneous reduction in star formation at the time of reionization in the star formation histories of local group dwarf galaxies.  Because galaxies well below the atomic cooling threshold remain self-shielded long after reionization, we expect that they will continue to form stars until their gas supply has been depleted due to lack of inflows or expulsion by SN feedback.

\end{itemize}

In this work, we have primarily focused on haloes in low-density environments in order to isolate the effects of radiation feedback on the properties of dwarf galaxies.  However, such low-mass haloes populate all different environments thus understanding how radiation feedback impacts these galaxies under a diverse set of conditions where other feedback modes are present (i.e., in the presence of a massive galaxy or AGN) remains to be determined.  Future simulations that focus on such processes can be combined with the results presented here to elucidate the physics that is responsible for quenching dwarf galaxies in all environments throughout the Universe.

\section*{Acknowledgements}
We thank the referee for their detailed and constructive review of the manuscript. HK thanks Brasenose College and the support of the Nicholas Kurti Junior Fellowship and the Beecroft Fellowship. Support by ERC Advanced Grant 320596 ``The Emergence of Structure during the Epoch of reionization'' is gratefully acknowledged.  TK was supported in part by the Yonsei University Future-leading Research Initiative (RMS2-2019-22-0216) and in part by the National Research Foundation of Korea (NRF-2017R1A5A1070354 and NRF-2018R1C1B5036146).  JR and JB acknowledge support from the ORAGE project from the Agence Nationale de la Recherche under grand ANR-14-CE33-0016-03.  TG acknowledges support from the European Research Council under grant agreement ERC-stg-757258 (TRIPLE).  The research of AS and JD is supported by Adrian Beecroft and STFC.  The results of this research have been achieved using the PRACE Research Infrastructure resource SuperMUC based in Garching, Germany

\appendix
\section{Tables of Stellar and Baryonic Content of Galaxies During the Epoch of Reionization}
\label{fitfunc}
In Tables~\ref{A1} and \ref{A2}, we list the evolution of the baryon fraction of galaxies ($\log_{10}({\rm M_b}/f_{\rm b}{\rm M_{vir}})$) and $f_{\rm bright}$, respectively, as a function of virial mass and redshift for the simulation that includes reionization.  A bin is only reported if it contains at least 10 systems.  Since $f_{\rm bright}$ saturates at 1 at a virial mass that is well resolved by our simulation, it is reasonable to extrapolate our relation to higher halo masses.  In contrast, there is a critical halo mass above which SN feedback dominates over radiation feedback.  This transition mass (${\rm M_{R/SN}}$) can be approximated using the following piecewise-linear function:
\begin{equation}
\log_{10}({\rm M_{R/SN}})=
\begin{cases} 
      -0.38(z-7.0) + 7.82 & z < 7.0 \\
      -0.05(z-7.0) + 7.82 & z \geq 7.0.
\end{cases}
\end{equation}
Thus if one wishes to model only the impact of radiation feedback, we recommend using only the values in our baryon fraction tables at masses less than ${\rm M_{R/SN}}$.

\begin{table*}
\centering
\begin{tabular}{lccccccc}
\hline
 &  \multicolumn{7}{|c|}{$\log_{10}({\rm M_{vir}/M_{\odot}})$}\\
Redshift & 7.0 &  7.33 &  7.66 &  8.0 &  8.33 &  8.66 &  9.0   \\
\hline
\hline
6.00 & $-1.78\pm0.36$ & $-1.43\pm0.48$ & $-0.70\pm0.45$ & $-0.33\pm0.31$ & $-0.32\pm0.32$ & $-0.35\pm0.20$ & $-0.33\pm0.18$  \\
6.08 & $-1.72\pm0.40$ & $-1.29\pm0.52$ & $-0.63\pm0.44$ & $-0.28\pm0.29$ & $-0.32\pm0.33$ & $-0.34\pm0.20$ & $-0.31\pm0.18$  \\
6.16 & $-1.63\pm0.44$ & $-1.17\pm0.55$ & $-0.56\pm0.45$ & $-0.26\pm0.29$ & $-0.31\pm0.33$ & $-0.31\pm0.19$ & $-0.29\pm0.18$  \\
6.24 & $-1.55\pm0.49$ & $-1.06\pm0.58$ & $-0.47\pm0.40$ & $-0.23\pm0.29$ & $-0.28\pm0.32$ & $-0.32\pm0.20$ & $-0.28\pm0.18$  \\
6.32 & $-1.44\pm0.55$ & $-0.95\pm0.59$ & $-0.42\pm0.43$ & $-0.23\pm0.30$ & $-0.27\pm0.32$ & $-0.33\pm0.20$ & $-0.28\pm0.19$  \\
6.35 & $-1.43\pm0.56$ & $-0.91\pm0.59$ & $-0.41\pm0.42$ & $-0.22\pm0.29$ & $-0.27\pm0.32$ & $-0.31\pm0.19$ & $-0.28\pm0.19$  \\
6.43 & $-1.33\pm0.61$ & $-0.81\pm0.60$ & $-0.37\pm0.41$ & $-0.21\pm0.29$ & $-0.25\pm0.30$ & $-0.29\pm0.19$ & $-0.31\pm0.18$  \\
6.52 & $-1.25\pm0.64$ & $-0.73\pm0.60$ & $-0.33\pm0.40$ & $-0.19\pm0.28$ & $-0.23\pm0.28$ & $-0.30\pm0.18$ & $-0.30\pm0.19$  \\
6.61 & $-1.15\pm0.66$ & $-0.67\pm0.61$ & $-0.29\pm0.36$ & $-0.18\pm0.26$ & $-0.22\pm0.28$ & $-0.30\pm0.18$ & $-0.31\pm0.21$  \\
6.71 & $-1.07\pm0.68$ & $-0.61\pm0.57$ & $-0.24\pm0.28$ & $-0.16\pm0.25$ & $-0.21\pm0.27$ & $-0.30\pm0.20$ & $-0.30\pm0.22$  \\
6.81 & $-1.00\pm0.67$ & $-0.55\pm0.54$ & $-0.22\pm0.26$ & $-0.16\pm0.25$ & $-0.22\pm0.27$ & $-0.30\pm0.19$ & $-0.28\pm0.19$  \\
6.91 & $-0.97\pm0.69$ & $-0.51\pm0.50$ & $-0.20\pm0.23$ & $-0.16\pm0.26$ & $-0.21\pm0.29$ & $-0.31\pm0.20$ & $-0.29\pm0.19$  \\
7.00 & $-0.92\pm0.69$ & $-0.46\pm0.47$ & $-0.18\pm0.22$ & $-0.15\pm0.26$ & $-0.23\pm0.28$ & $-0.35\pm0.20$  \\
7.08 & $-0.86\pm0.66$ & $-0.43\pm0.45$ & $-0.16\pm0.21$ & $-0.15\pm0.27$ & $-0.22\pm0.27$ & $-0.33\pm0.21$  \\
7.19 & $-0.80\pm0.65$ & $-0.40\pm0.42$ & $-0.14\pm0.18$ & $-0.15\pm0.28$ & $-0.23\pm0.26$ & $-0.30\pm0.22$  \\
7.31 & $-0.75\pm0.63$ & $-0.37\pm0.39$ & $-0.13\pm0.18$ & $-0.15\pm0.28$ & $-0.23\pm0.25$ & $-0.32\pm0.23$  \\
7.42 & $-0.72\pm0.61$ & $-0.35\pm0.37$ & $-0.12\pm0.16$ & $-0.14\pm0.28$ & $-0.23\pm0.25$ & $-0.32\pm0.23$ & $-0.37\pm0.14$  \\
7.55 & $-0.66\pm0.58$ & $-0.33\pm0.36$ & $-0.12\pm0.15$ & $-0.14\pm0.27$ & $-0.23\pm0.25$ & $-0.32\pm0.22$ & $-0.38\pm0.14$  \\
7.67 & $-0.60\pm0.57$ & $-0.29\pm0.32$ & $-0.10\pm0.13$ & $-0.14\pm0.27$ & $-0.21\pm0.26$ & $-0.30\pm0.19$ & $-0.35\pm0.14$  \\
7.80 & $-0.56\pm0.52$ & $-0.27\pm0.30$ & $-0.09\pm0.13$ & $-0.15\pm0.28$ & $-0.23\pm0.27$ & $-0.28\pm0.20$ & $-0.33\pm0.12$  \\
7.89 & $-0.54\pm0.51$ & $-0.26\pm0.30$ & $-0.09\pm0.11$ & $-0.14\pm0.27$ & $-0.23\pm0.27$ & $-0.26\pm0.19$  \\
8.03 & $-0.52\pm0.50$ & $-0.24\pm0.27$ & $-0.09\pm0.13$ & $-0.13\pm0.28$ & $-0.22\pm0.27$ & $-0.29\pm0.22$  \\
8.18 & $-0.48\pm0.46$ & $-0.22\pm0.24$ & $-0.09\pm0.13$ & $-0.13\pm0.27$ & $-0.19\pm0.24$ & $-0.34\pm0.25$  \\
8.33 & $-0.46\pm0.50$ & $-0.20\pm0.20$ & $-0.08\pm0.13$ & $-0.11\pm0.25$ & $-0.21\pm0.24$ & $-0.33\pm0.21$  \\
8.49 & $-0.45\pm0.50$ & $-0.19\pm0.22$ & $-0.08\pm0.15$ & $-0.11\pm0.23$ & $-0.21\pm0.28$ & $-0.37\pm0.23$  \\
8.65 & $-0.39\pm0.44$ & $-0.18\pm0.20$ & $-0.07\pm0.13$ & $-0.09\pm0.20$ & $-0.20\pm0.24$ & $-0.38\pm0.26$  \\
8.83 & $-0.32\pm0.33$ & $-0.16\pm0.18$ & $-0.07\pm0.13$ & $-0.10\pm0.21$ & $-0.20\pm0.22$ & $-0.39\pm0.25$  \\
9.01 & $-0.30\pm0.31$ & $-0.15\pm0.18$ & $-0.07\pm0.13$ & $-0.12\pm0.25$ & $-0.22\pm0.23$ & $-0.47\pm0.29$  \\
9.20 & $-0.28\pm0.29$ & $-0.14\pm0.18$ & $-0.07\pm0.14$ & $-0.12\pm0.24$ & $-0.26\pm0.26$ & $-0.49\pm0.25$  \\
9.39 & $-0.25\pm0.28$ & $-0.12\pm0.11$ & $-0.07\pm0.13$ & $-0.11\pm0.24$ & $-0.25\pm0.22$  \\
9.60 & $-0.22\pm0.22$ & $-0.11\pm0.10$ & $-0.06\pm0.13$ & $-0.10\pm0.23$ & $-0.22\pm0.18$ & $-0.48\pm0.25$  \\
9.82 & $-0.21\pm0.25$ & $-0.11\pm0.09$ & $-0.06\pm0.14$ & $-0.12\pm0.25$ & $-0.22\pm0.20$  \\
10.05 & $-0.22\pm0.26$ & $-0.10\pm0.09$ & $-0.07\pm0.14$ & $-0.15\pm0.27$ & $-0.24\pm0.19$  \\
10.29 & $-0.19\pm0.16$ & $-0.09\pm0.08$ & $-0.06\pm0.14$ & $-0.16\pm0.25$ & $-0.26\pm0.21$  \\
10.46 & $-0.18\pm0.15$ & $-0.09\pm0.07$ & $-0.06\pm0.11$ & $-0.19\pm0.26$ & $-0.26\pm0.22$  \\
10.73 & $-0.16\pm0.13$ & $-0.08\pm0.06$ & $-0.06\pm0.12$ & $-0.16\pm0.20$ & $-0.26\pm0.23$  \\
11.01 & $-0.14\pm0.11$ & $-0.08\pm0.06$ & $-0.06\pm0.11$ & $-0.17\pm0.20$ & $-0.25\pm0.23$  \\
11.31 & $-0.13\pm0.09$ & $-0.07\pm0.06$ & $-0.05\pm0.12$ & $-0.20\pm0.26$ & $-0.29\pm0.21$  \\
11.63 & $-0.13\pm0.09$ & $-0.07\pm0.05$ & $-0.07\pm0.15$ & $-0.20\pm0.27$  \\
11.97 & $-0.13\pm0.09$ & $-0.07\pm0.05$ & $-0.10\pm0.19$ & $-0.15\pm0.22$ & $-0.37\pm0.14$  \\
12.33 & $-0.10\pm0.07$ & $-0.06\pm0.05$ & $-0.09\pm0.18$ & $-0.14\pm0.20$ & $-0.28\pm0.16$  \\
12.72 & $-0.10\pm0.07$ & $-0.06\pm0.04$ & $-0.07\pm0.15$ & $-0.17\pm0.19$  \\
13.15 & $-0.11\pm0.07$ & $-0.06\pm0.04$ & $-0.08\pm0.14$ & $-0.12\pm0.13$  \\
13.60 & $-0.10\pm0.06$ & $-0.07\pm0.04$ & $-0.07\pm0.11$ & $-0.07\pm0.10$  \\
14.09 & $-0.10\pm0.05$ & $-0.06\pm0.04$ & $-0.07\pm0.07$  \\
14.63 & $-0.10\pm0.05$ & $-0.05\pm0.03$ & $-0.04\pm0.03$  \\
15.22 & $-0.10\pm0.05$ & $-0.05\pm0.04$ & $-0.03\pm0.02$  \\
15.87 & $-0.09\pm0.04$ & $-0.06\pm0.08$  \\
16.58 & $-0.07\pm0.05$ & $-0.08\pm0.10$  \\
17.38 & $-0.08\pm0.04$  \\
\hline
\end{tabular}
\caption{Baryon fractions as a function of virial mass and redshift.  Quantities represent $\log_{10}({\rm M_b}/f_{\rm b}{\rm M_{vir}})$ and the quoted uncertainty represents the $1\sigma$ scatter at fixed virial mass.}
\label{A1}
\end{table*}%

\begin{table*}
\centering
\begin{tabular}{lccccccc}
\hline
 &  \multicolumn{7}{|c|}{$\log_{10}({\rm M_{vir}/M_{\odot}})$}\\
Redshift & 7.0 &  7.33 &  7.66 &  8.0 &  8.33 &  8.66 &  9.0   \\
\hline
\hline
6.00 & $0.000\pm0.00$ & $0.000\pm0.03$ & $0.030\pm0.18$ & $0.490\pm0.50$ & $0.880\pm0.32$ & $1.000\pm0.00$ & $1.000\pm0.00$  \\
6.08 & $0.000\pm0.00$ & $0.000\pm0.03$ & $0.030\pm0.18$ & $0.500\pm0.50$ & $0.880\pm0.33$ & $1.000\pm0.00$ & $1.000\pm0.00$  \\
6.16 & $0.000\pm0.00$ & $0.000\pm0.03$ & $0.030\pm0.18$ & $0.510\pm0.50$ & $0.880\pm0.33$ & $1.000\pm0.00$ & $1.000\pm0.00$  \\
6.24 & $0.000\pm0.00$ & $0.000\pm0.04$ & $0.040\pm0.19$ & $0.530\pm0.50$ & $0.850\pm0.35$ & $1.000\pm0.00$ & $1.000\pm0.00$  \\
6.32 & $0.000\pm0.00$ & $0.000\pm0.04$ & $0.030\pm0.16$ & $0.500\pm0.50$ & $0.880\pm0.33$ & $1.000\pm0.00$ & $1.000\pm0.00$  \\
6.35 & $0.000\pm0.00$ & $0.000\pm0.04$ & $0.030\pm0.17$ & $0.510\pm0.50$ & $0.870\pm0.33$ & $0.970\pm0.17$ & $1.000\pm0.00$  \\
6.43 & $0.000\pm0.00$ & $0.000\pm0.00$ & $0.040\pm0.19$ & $0.520\pm0.50$ & $0.880\pm0.33$ & $0.970\pm0.18$ & $1.000\pm0.00$  \\
6.52 & $0.000\pm0.00$ & $0.000\pm0.00$ & $0.050\pm0.21$ & $0.540\pm0.50$ & $0.890\pm0.31$ & $1.000\pm0.00$ & $1.000\pm0.00$  \\
6.61 & $0.000\pm0.00$ & $0.000\pm0.00$ & $0.040\pm0.21$ & $0.530\pm0.50$ & $0.890\pm0.31$ & $1.000\pm0.00$ & $1.000\pm0.00$  \\
6.71 & $0.000\pm0.00$ & $0.000\pm0.00$ & $0.040\pm0.20$ & $0.530\pm0.50$ & $0.920\pm0.26$ & $1.000\pm0.00$ & $1.000\pm0.00$  \\
6.81 & $0.000\pm0.00$ & $0.000\pm0.00$ & $0.040\pm0.20$ & $0.540\pm0.50$ & $0.940\pm0.24$ & $1.000\pm0.00$ & $1.000\pm0.00$  \\
6.91 & $0.000\pm0.00$ & $0.000\pm0.00$ & $0.040\pm0.20$ & $0.590\pm0.49$ & $0.930\pm0.26$ & $0.970\pm0.16$ & $1.000\pm0.00$  \\
7.00 & $0.000\pm0.00$ & $0.000\pm0.00$ & $0.040\pm0.19$ & $0.580\pm0.49$ & $0.930\pm0.26$ & $1.000\pm0.00$  \\
7.08 & $0.000\pm0.00$ & $0.000\pm0.00$ & $0.040\pm0.20$ & $0.610\pm0.49$ & $0.890\pm0.31$ & $1.000\pm0.00$  \\
7.19 & $0.000\pm0.00$ & $0.000\pm0.00$ & $0.040\pm0.20$ & $0.600\pm0.49$ & $0.870\pm0.33$ & $1.000\pm0.00$  \\
7.31 & $0.000\pm0.00$ & $0.000\pm0.00$ & $0.060\pm0.23$ & $0.650\pm0.48$ & $0.920\pm0.27$ & $1.000\pm0.00$  \\
7.42 & $0.000\pm0.00$ & $0.000\pm0.00$ & $0.050\pm0.22$ & $0.680\pm0.47$ & $0.950\pm0.22$ & $1.000\pm0.00$ & $1.000\pm0.00$  \\
7.55 & $0.000\pm0.00$ & $0.000\pm0.00$ & $0.070\pm0.25$ & $0.670\pm0.47$ & $0.930\pm0.26$ & $1.000\pm0.00$ & $1.000\pm0.00$  \\
7.67 & $0.000\pm0.00$ & $0.000\pm0.00$ & $0.070\pm0.25$ & $0.630\pm0.48$ & $0.910\pm0.29$ & $1.000\pm0.00$ & $1.000\pm0.00$  \\
7.80 & $0.000\pm0.00$ & $0.000\pm0.00$ & $0.060\pm0.24$ & $0.620\pm0.48$ & $0.940\pm0.23$ & $1.000\pm0.00$ & $1.000\pm0.00$  \\
7.89 & $0.000\pm0.00$ & $0.000\pm0.00$ & $0.060\pm0.24$ & $0.640\pm0.48$ & $0.960\pm0.21$ & $1.000\pm0.00$  \\
8.03 & $0.000\pm0.00$ & $0.000\pm0.00$ & $0.070\pm0.26$ & $0.700\pm0.46$ & $0.970\pm0.18$ & $1.000\pm0.00$  \\
8.18 & $0.000\pm0.00$ & $0.000\pm0.00$ & $0.080\pm0.27$ & $0.720\pm0.45$ & $0.930\pm0.25$ & $1.000\pm0.00$  \\
8.33 & $0.000\pm0.00$ & $0.000\pm0.00$ & $0.090\pm0.28$ & $0.740\pm0.44$ & $0.950\pm0.22$ & $1.000\pm0.00$  \\
8.49 & $0.000\pm0.00$ & $0.000\pm0.00$ & $0.080\pm0.28$ & $0.730\pm0.44$ & $0.960\pm0.19$ & $1.000\pm0.00$  \\
8.65 & $0.000\pm0.00$ & $0.000\pm0.00$ & $0.090\pm0.29$ & $0.710\pm0.46$ & $0.960\pm0.19$ & $1.000\pm0.00$  \\
8.83 & $0.000\pm0.00$ & $0.000\pm0.00$ & $0.110\pm0.32$ & $0.710\pm0.45$ & $0.980\pm0.13$ & $1.000\pm0.00$  \\
9.01 & $0.000\pm0.00$ & $0.000\pm0.00$ & $0.120\pm0.33$ & $0.710\pm0.45$ & $0.980\pm0.13$ & $1.000\pm0.00$  \\
9.20 & $0.000\pm0.00$ & $0.000\pm0.00$ & $0.130\pm0.34$ & $0.760\pm0.43$ & $0.950\pm0.21$ & $1.000\pm0.00$  \\
9.39 & $0.000\pm0.00$ & $0.000\pm0.00$ & $0.140\pm0.35$ & $0.750\pm0.44$ & $1.000\pm0.00$  \\
9.60 & $0.000\pm0.00$ & $0.000\pm0.00$ & $0.150\pm0.36$ & $0.720\pm0.45$ & $0.970\pm0.17$ & $1.000\pm0.00$  \\
9.82 & $0.000\pm0.00$ & $0.000\pm0.00$ & $0.190\pm0.39$ & $0.760\pm0.43$ & $0.960\pm0.19$  \\
10.05 & $0.000\pm0.00$ & $0.000\pm0.00$ & $0.220\pm0.41$ & $0.770\pm0.42$ & $0.960\pm0.19$  \\
10.29 & $0.000\pm0.00$ & $0.000\pm0.00$ & $0.260\pm0.44$ & $0.760\pm0.43$ & $0.960\pm0.19$  \\
10.46 & $0.000\pm0.00$ & $0.000\pm0.00$ & $0.250\pm0.44$ & $0.850\pm0.36$ & $1.000\pm0.00$  \\
10.73 & $0.000\pm0.00$ & $0.000\pm0.00$ & $0.290\pm0.45$ & $0.920\pm0.27$ & $1.000\pm0.00$  \\
11.01 & $0.000\pm0.00$ & $0.000\pm0.00$ & $0.300\pm0.46$ & $0.870\pm0.34$ & $1.000\pm0.00$  \\
11.31 & $0.000\pm0.00$ & $0.000\pm0.00$ & $0.310\pm0.46$ & $0.900\pm0.30$ & $1.000\pm0.00$  \\
11.63 & $0.000\pm0.00$ & $0.000\pm0.00$ & $0.370\pm0.48$ & $0.950\pm0.21$  \\
11.97 & $0.000\pm0.00$ & $0.010\pm0.10$ & $0.460\pm0.50$ & $1.000\pm0.00$ & $1.000\pm0.00$  \\
12.33 & $0.000\pm0.00$ & $0.010\pm0.08$ & $0.480\pm0.50$ & $1.000\pm0.00$ & $1.000\pm0.00$  \\
12.72 & $0.000\pm0.00$ & $0.010\pm0.07$ & $0.480\pm0.50$ & $0.950\pm0.22$  \\
13.15 & $0.000\pm0.00$ & $0.010\pm0.09$ & $0.520\pm0.50$ & $0.860\pm0.35$  \\
13.60 & $0.000\pm0.00$ & $0.010\pm0.10$ & $0.710\pm0.45$ & $0.930\pm0.26$  \\
14.09 & $0.000\pm0.00$ & $0.070\pm0.25$ & $0.680\pm0.46$  \\
14.63 & $0.000\pm0.00$ & $0.070\pm0.25$ & $0.630\pm0.48$  \\
15.22 & $0.000\pm0.00$ & $0.080\pm0.27$ & $0.460\pm0.50$  \\
15.87 & $0.000\pm0.00$ & $0.120\pm0.32$  \\
16.58 & $0.000\pm0.00$ & $0.170\pm0.37$  \\
17.38 & $0.000\pm0.00$  \\
\hline
\end{tabular}
\caption{$f_{\rm bright}$ as a function of virial mass and redshift.  Quantities represent the fraction of haloes at a given virial mass and redshift that form at least 1000M$_{\odot}$  in stars and the quoted uncertainty represents the $1\sigma$ scatter at fixed virial mass.}
\label{A2}
\end{table*}%

\bibliographystyle{mnras}
\bibliography{mnras_guide} 

\bsp	
\label{lastpage}
\end{document}